\documentclass[aps,prb,twocolumn,superscriptaddress,reprint,nofootinbib]{revtex4-2}
\usepackage{graphicx,color}
\usepackage{amsmath}
\usepackage{amssymb}

\usepackage[utf8]{inputenc}
\usepackage[english]{babel}
\graphicspath{{images/}}

\usepackage{float}
\usepackage{physics}
\usepackage{indentfirst}
\usepackage{upgreek}

\usepackage{lipsum}

\usepackage{xcolor}

\usepackage{titlesec}

\usepackage{hyperref}
\hypersetup{
	colorlinks   = true,    
	urlcolor     = blue,    
	linkcolor    = blue,    
	citecolor    = blue      
}

\usepackage{blindtext}

\usepackage[font=small,labelfont=bf,labelsep=period]{caption}
\usepackage[singlelinecheck=false,justification=raggedright]{caption}

\begin{document}
	
	
	\title{Bulk and surface electron scattering in disordered Bi$_{2}$Te$_{3}$ probed by quasiparticle interference}
	
	
	\author{V. Nagorkin}
	\email[]{vladislav.nagorkin@uni-wuppertal.de}
	\affiliation{Leibniz Institute for Solid State and Materials Research, IFW Dresden, 01069 Dresden, Germany}
	\affiliation{Fakult{\"a}t f{\"u}r Mathematik und Naturwissenschaften, Bergische Universit{\"a}t Wuppertal, 42097 Wuppertal, Germany}
	
	\author{S. Schimmel}
	\affiliation{Leibniz Institute for Solid State and Materials Research, IFW Dresden, 01069 Dresden, Germany}
	\affiliation{Fakult{\"a}t f{\"u}r Mathematik und Naturwissenschaften, Bergische Universit{\"a}t Wuppertal, 42097 Wuppertal, Germany}
	
	\author{P. Gebauer}
	\affiliation{Faculty of Chemistry and Food Chemistry, Technische Universit{\"a}t Dresden, 01069 Dresden, Germany}
	
	\author{A. Isaeva}
	\affiliation{Leibniz Institute for Solid State and Materials Research, IFW Dresden, 01069 Dresden, Germany}
	\affiliation{Van der Waals–Zeeman Institute, Department of Physics and Astronomy, University of Amsterdam, Amsterdam 1098 XH, The Netherlands}
	\affiliation{Department of Physics, Technische Universit{\"a}t Dortmund, 44227 Dortmund, Germany}
	\affiliation{Research Center "Future Energy Materials and Systems", UA Ruhr}
	
	\author{D. Baumann}
	\affiliation{Leibniz Institute for Solid State and Materials Research, IFW Dresden, 01069 Dresden, Germany}	
	
	\author{A. Koitzsch}
	\affiliation{Leibniz Institute for Solid State and Materials Research, IFW Dresden, 01069 Dresden, Germany}
	
	\author{B. B{\"u}chner}
	\affiliation{Leibniz Institute for Solid State and Materials Research, IFW Dresden, 01069 Dresden, Germany}
	\affiliation{Faculty of Physics, Technische Universit{\"a}t Dresden, 01069 Dresden, Germany}
	\affiliation{Institute of Solid State and Materials Physics and W{\"u}rzburg-Dresden Cluster of Excellence ct.qmat, Technische Universit{\"a}t Dresden, 01062 Dresden, Germany}
	\affiliation{Center for Transport and Devices, Technische Universit{\"a}t Dresden, 01069 Dresden, Germany}
	
	\author{C. Hess}
	\affiliation{Leibniz Institute for Solid State and Materials Research, IFW Dresden, 01069 Dresden, Germany}
	\affiliation{Fakult{\"a}t f{\"u}r Mathematik und Naturwissenschaften,
	Bergische Universit{\"a}t Wuppertal, 42097 Wuppertal, Germany}
	\affiliation{Institute of Solid State and Materials Physics and W{\"u}rzburg-Dresden Cluster of Excellence ct.qmat, Technische Universit{\"a}t Dresden, 01062 Dresden, Germany}
	\affiliation{Center for Transport and Devices, Technische Universit{\"a}t Dresden, 01069 Dresden, Germany}

	\date{\today}
	
	\begin{abstract}
		
		We investigated the electronic properties of the topological insulator Bi$_{2}$Te$_{3}$ by scanning tunneling microscopy and spectroscopy at low temperature. We obtained high-resolution quasiparticle interference data of the topological surface Dirac electrons at different energies. Spin-selective joint density of states calculations were performed for surface and bulk electronic states to interpret the observed quasiparticle interference data. The topological properties of our crystals are demonstrated by the absence of backscattering along with the linear energy dispersion of the dominant scattering vector. In addition, we detect non-dispersive scattering modes which we associate with bulk-surface scattering and, thus, allow an approximate identification of the bulk energy gap range based on our quasiparticle interference data. Measurements of differential conductance maps 
		in magnetic fields up to 15 T have been carried out, but no strong modifications could be observed.
				
	\end{abstract}

	\maketitle

	\section{Introduction}
	
	In recent years, three-dimensional topological insulators have been at the forefront of research in condensed matter physics due to their not only fundamentally new physical properties, but also practical importance for potential applications in spintronics and quantum computing \cite{PhysRevLett.98.106803, PhysRevB.75.121306, Moore2009, Zhang2009, Xia2009, doi:10.1126/science.1173034,RevModPhys.82.3045}. The key features of these materials are their gapless Dirac surface state with spin-momentum locking of the Dirac electrons, resulting in the suppression of backscattering. In addition, some compounds like Bi$_{2}$Te$_{3}$ show a peculiar evolution of the shape of the surface state's constant-energy contours (CECs). In particular, the circular shape of the CEC in Bi$_{2}$Te$_{3}$ transforms into a hexagon which further becomes a concave hexagram due to the hexagonal warping effect upon increasing the energy starting from the Dirac point \cite{PhysRevLett.103.266801, 2009, doi:10.1126/science.1173034}. Unlike the ideal shape of the Dirac cone, the warping in Bi$_{2}$Te$_{3}$ leads to an enhanced quasiparticle interference (QPI) \cite{PhysRevLett.103.266801, 2009}. The QPI results from elastic electron scattering off impurities or crystal defects and is manifested in spatial modulations of the local density of states (LDOS), which can be probed by scanning tunneling microscopy and spectroscopy (STM/STS). The QPI typically is detected in differential conductance ($dI/dU$) maps, and the Fourier transforms of such data (FT-$dI/dU$) provide unique information on scattering processes and the electronic structure in reciprocal space \cite{Sprunger1997, PhysRevB.57.R6858}. To date, the analysis of QPI has been established as a powerful method for studying various electronic systems, for instance, high-temperature superconductors \cite{Hoffman2002}, heavy-fermion materials \cite{Schmidt2010, Aynajian2012}, Weyl semimetals \cite{Zheng2016, Inoue2016} as well as topological insulators \cite{Roushan2009, PhysRevLett.103.266803, PhysRevLett.104.016401}. For the latter, the absence of backscattering due to the spin-momentum locking was clearly demonstrated in those studies.
		
	Previous STM/STS and QPI investigations provided meaningful insights about the electronic properties of the prototypical topological insulator Bi$_{2}$Te$_{3}$ \cite{PhysRevB.69.085313, PhysRevLett.103.266803, PhysRevLett.104.016401, PhysRevLett.109.166407,PhysRevB.88.161407, sessi2014signatures, https://doi.org/10.1002/adma.201504771, netsou2020identifying, Stolyarov2021}. Building on this, we conducted low-temperature STM/STS measurements on single crystals of Bi$_{2}$Te$_{3}$ with the main focus on scattering processes revealed by QPI. Here, we report on the interplay between bulk and surface electron scattering and the influence of the intricate spin texture with the out-of-plane spin component \cite{PhysRevLett.103.266801, Hsieh2009, PhysRevLett.106.216803} on possible scattering processes. High-resolution Fourier-transformed QPI (FT-QPI) patterns were measured in a wide energy range [$-600$; 300] meV to characterize the material's dispersive electronic band structure. We interpret our data by comparing them with spin-selective joint density of states (JDOS) calculations. Within this approach, the analysis reveals the hallmark property of topological surface states: the absence of the backscattering and the linear energy dispersion. Furthermore, interestingly, we detected non-dispersive scattering modes in this compound, and their origin can be related to scattering involving both surface and bulk states. In addition, we have measured QPI in high magnetic fields up to 15 T. The absence of significant changes in the QPI patterns with respect to the zero-field data, in particular, suppression of backscattering corroborates that the	magnetic field does not lift the topological protection of the surface state.  
	
	\section{Experiment}
	
	Single crystals of Bi$_{2}$Te$_{3}$ were obtained by chemical vapor transport technique. A mixture of the elements was loaded into a silica ampoule together with transport agent (GdCl$_{3}$), sealed off under dynamic vacuum \\(p $\leq 1 \times 10^{-3}$ mbar), and put into a two-zone furnace into a temperature gradient 550$^{\circ}$C and 450$^{\circ}$C. After several days, large hexagonal plates of Bi$_{2}$Te$_{3}$ have formed in the colder end of the ampoule.
	Chemical composition and crystal structure were verified by semi-quantitative energy-dispersive X-ray spectroscopy (SU8020 (Hitachi) scanning transmission microsope, 20 kV acceleration voltage) and powder X-ray diffraction analysis (X’Pert Pro diffractometer (PANalytical), Bragg–Brentano geometry, curved Ge(111) monochromator, Cu-K$\alpha$1 radiation (${\lambda}$ = 1.5406 \AA{})), respectively. Prior to the STM measurements, the single crystals of Bi$_{2}$Te$_{3}$ were characterized by X-ray photoelectron spectroscopy which excluded possible extrinsic doping of the crystals, confirmed no deviation from the stoichiometry of this compound together with possible formation of surface bismuth oxides due to air exposure of the crystals (see more details in Fig.~\ref{fig:Fig.13} in Appendix). 
	
	The crystals were then cleaved in cryogenic vacuum at a base temperature of about 7 K just before the measurements at the same temperature in a home-built scanning tunneling microscope \cite{Schlegel2014} powered by a Nanonis Specs STM-controller \cite{Specs}. The $dI/dU$-signal was measured by use of an external SR 830 lock-in amplifier with $U_{mod}$ = (2$-$10) mV RMS and $f_{mod} = 0.667$ kHz. Magnetic field measurements were carried out with the use of a magnet cryostat with a maximal field $B = 15$ T. The STM tips were prepared by a standard electrochemical etching with a subsequent annealing treatment by electron beam heating. The STM data were analyzed with the WSxM \cite{Horcas2007} software and dedicated Python program codes. All the FT-QPI data were symmetrized according to the six-fold symmetric crystal structure in order to achieve better signal-to-noise ratio.  
	
	\section{Results and discussion}
	
	\subsection{STM measurements at zero-field}
	\label{subsec:STM measurements at zero-field }		

\begin{figure}[H]
		\centering
		\includegraphics[scale=0.48]{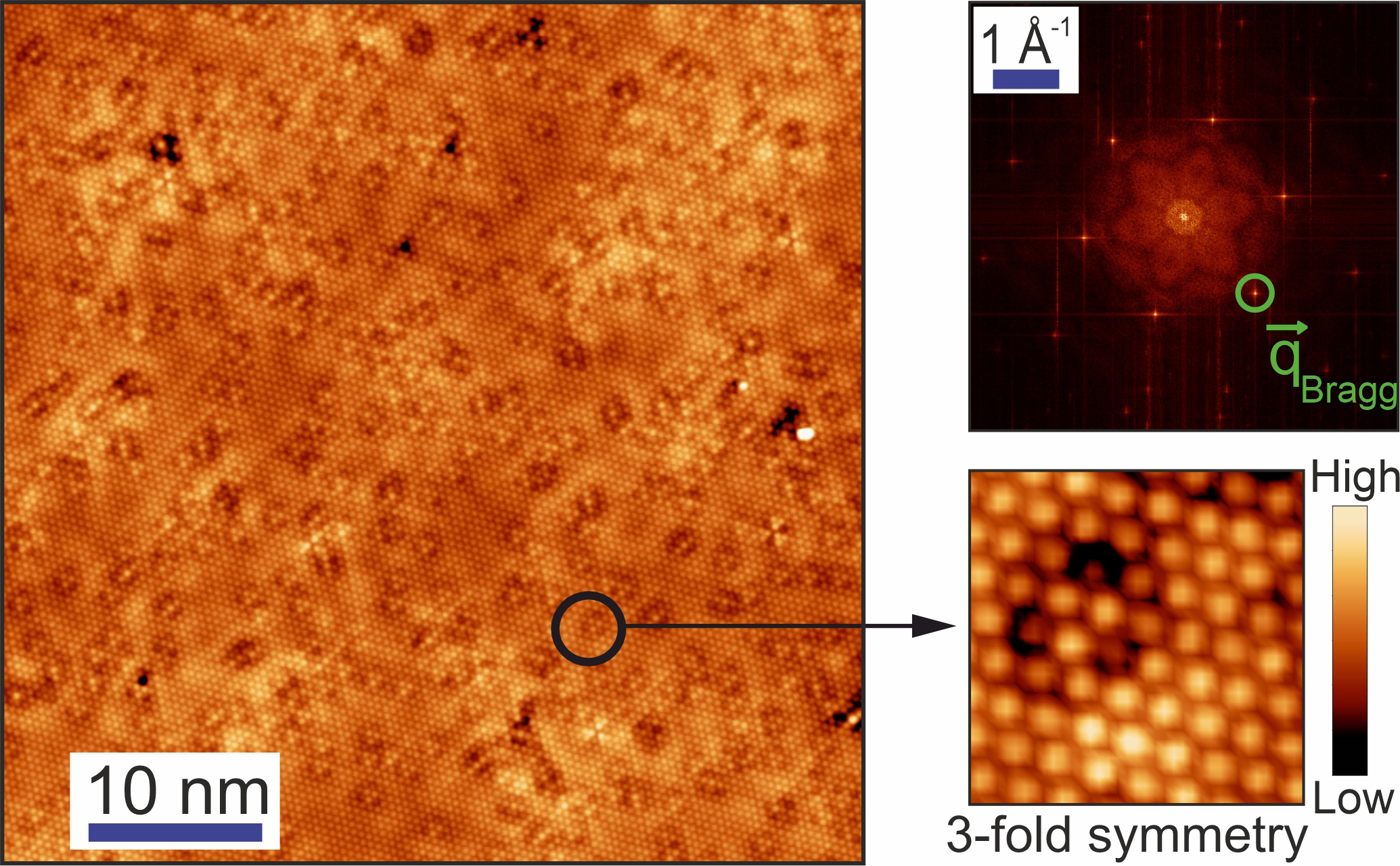}
		\caption{Representative atomically corrugated topography of Bi$_{2}$Te$_{3}$ measured in a field of view of 50 $\times$ 50 nm$^{2}$ at $U_{bias} = 100$ mV, $I_{T} = 100$ pA, $T = 7$ K with the FFT at the upper right corner. The FFT shows distinct Bragg reflection spots (one of them is marked with the green circle) indicating the material's hexagonal lattice and an additional six-fold symmetric pattern inside. The inner long-wavelength six-fold symmetric structure includes the energy integrated QPI and structures coming from the geometry of features in the topography. The dominant defect type is shown in an enlarged view at the bottom right corner.}
		\label{fig:Fig.1}
	\end{figure}

	Fig.~\ref{fig:Fig.1} shows a representative topographic map of the freshly cleaved sample surface in a 50 nm $\times$ 50 nm field of view. The data reveal the atomically resolved	surface exhibiting numerous defects. The disorder of the material can be recognized by the presence of the dominant defect type which occurs in about 2 \% of the unit cells. These defects are conceivably assigned to Bi substitutions of Te (or vacancies of Te) atomic sites in the third (Te) atomic layer (the middle one) inside the topmost quintuple Te-Bi-Te-Bi-Te layer \cite{netsou2020identifying}. Note that the most abundant defects exhibit a three-fold symmetric appearance (Fig.~\ref{fig:Fig.1}) suggestive of a three-fold symmetric scattering potential \cite{Beidenkopf2011, PhysRevB.88.161407, Stolyarov2021}. In section~\ref{subsec:QPI analysis }, the latter and its influence on the scattering properties and, hence, on the QPI patterns will be discussed further.
	
\begin{figure}[H]
		\centering
		\includegraphics[scale=0.52]{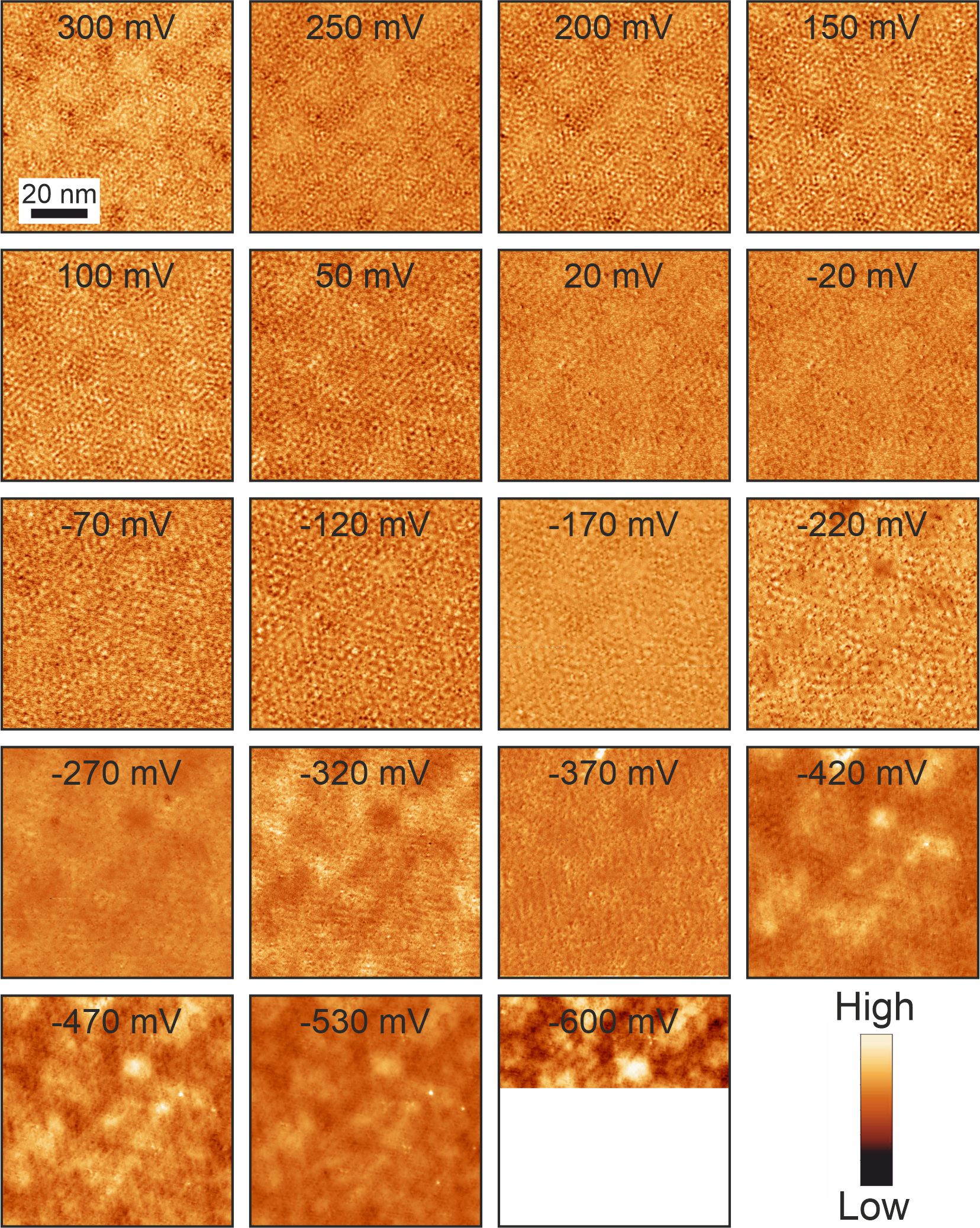}
		\caption{Series of the $dI/dU$ maps measured at the energies from 300 mV to $-600$ mV on the same surface area. The scalebar is the same for each of the maps.}
		\label{fig:Fig.2}
	\end{figure}

\begin{figure}[H]
		\centering
		\includegraphics[scale=0.43]{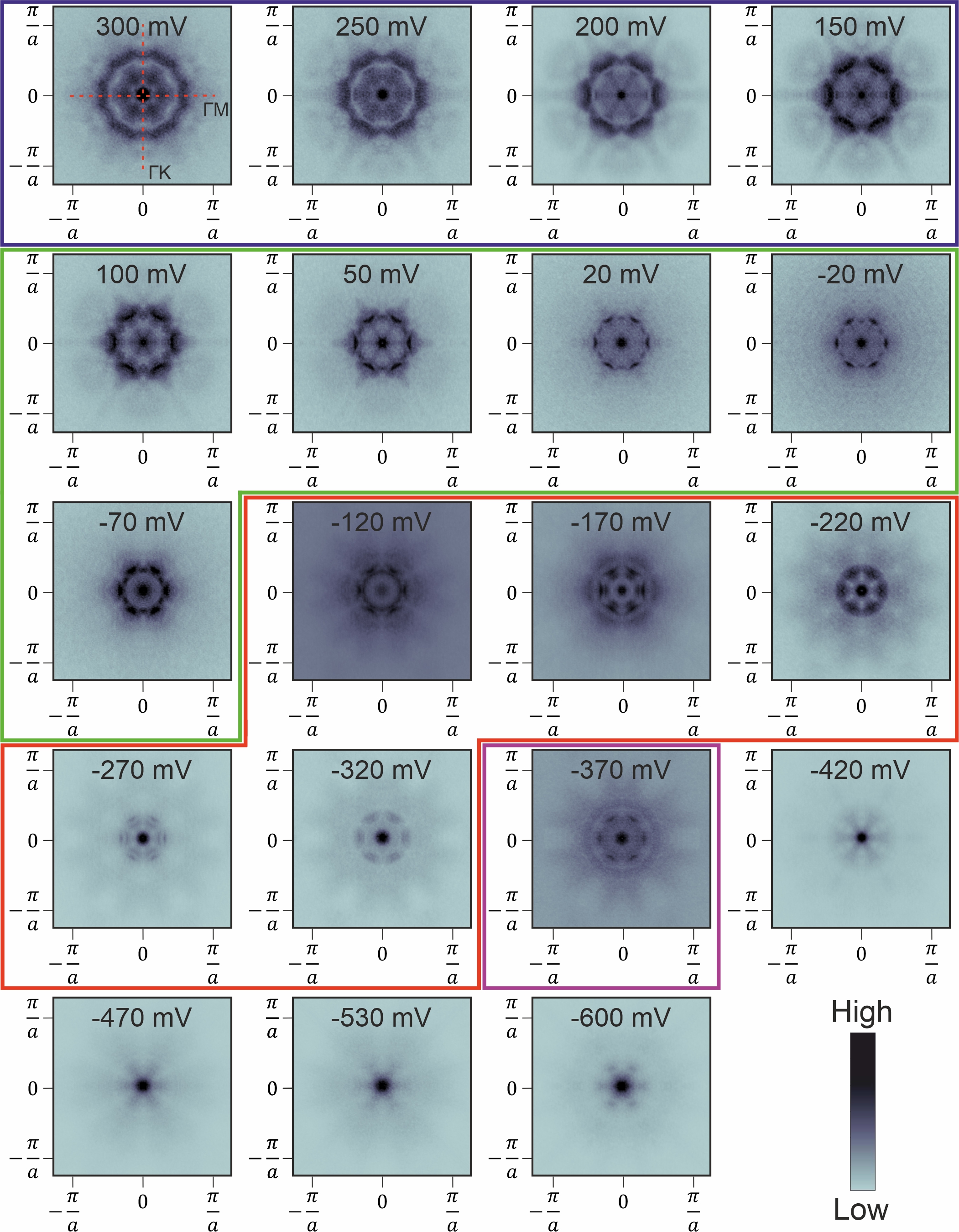}
		\caption{Series of the symmetrized FT-QPI patterns measured at the bias voltages from 300 mV to $-600$ mV on the same surface area. The maps surrounded by blue, green, red and magenta frames correspond to different types of patterns.}
		\label{fig:Fig.3}
	\end{figure}

\begin{figure}[H]
		\centering
		\includegraphics[scale=0.55]{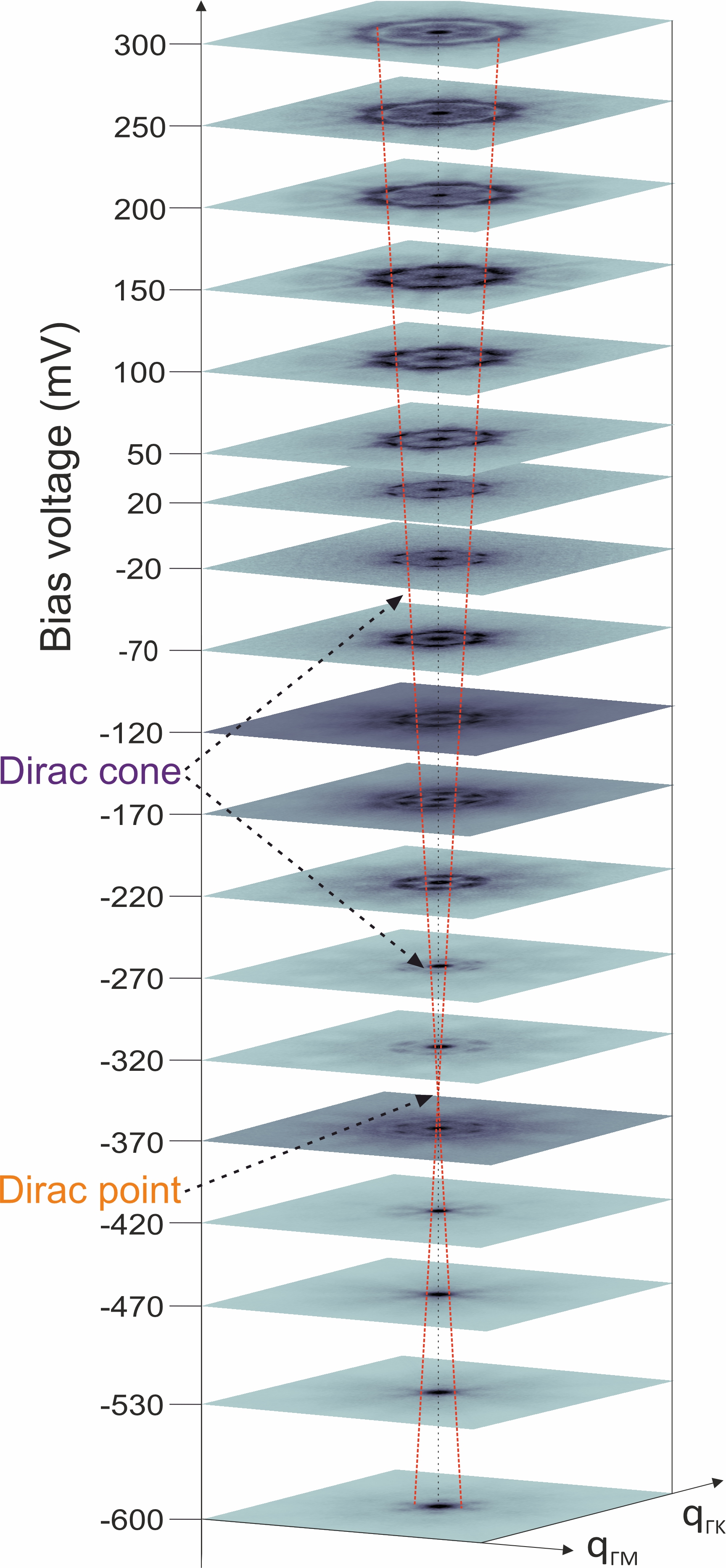}
		\caption{Stack of the FT-QPI patterns measured at $B = 0$ T. High-symmetry directions ${\Gamma}$-M and ${\Gamma}$-K are labelled as q$\textsubscript{${\Gamma}$M}$ and q$\textsubscript{${\Gamma}$K}$, respectively.}
		\label{fig:Fig.4}
	\end{figure}

	The $dI/dU$ maps were measured from $-600$ mV to 300 mV on the same surface area of 80 nm $\times$ 80 nm size and are presented with corresponding fast Fourier transforms (FFTs) in Figs.~\ref{fig:Fig.2}{\color{blue}} and~\ref{fig:Fig.3}{\color{blue}}, respectively. Note that the FT-QPI patterns acquired from maps measured on different samples share the characteristic features at respective energies (the data is supported by Figs.~\ref{fig:Fig.14} and~\ref{fig:Fig.15}), thus, the data are sample independent within the batch under investigation. Our data clearly demonstrate	a continuous increase in the modulation period in real space with decreasing the bias voltage from 300 mV down to $-70$ mV. At lower energies the patterns become more diffuse, and one has to rely on the FT-QPI data in order to infer the energy dependence (see Fig.~\ref{fig:Fig.3}{\color{blue}}). For better illustration we plot these data in Fig.~\ref{fig:Fig.4} as a stack, where we sketch the linear dispersion with a node at about $-355$ meV with the red dotted lines (see analysis below). 
	
\begin{figure*}[]
		\centering
		\includegraphics[width=0.98\textwidth]{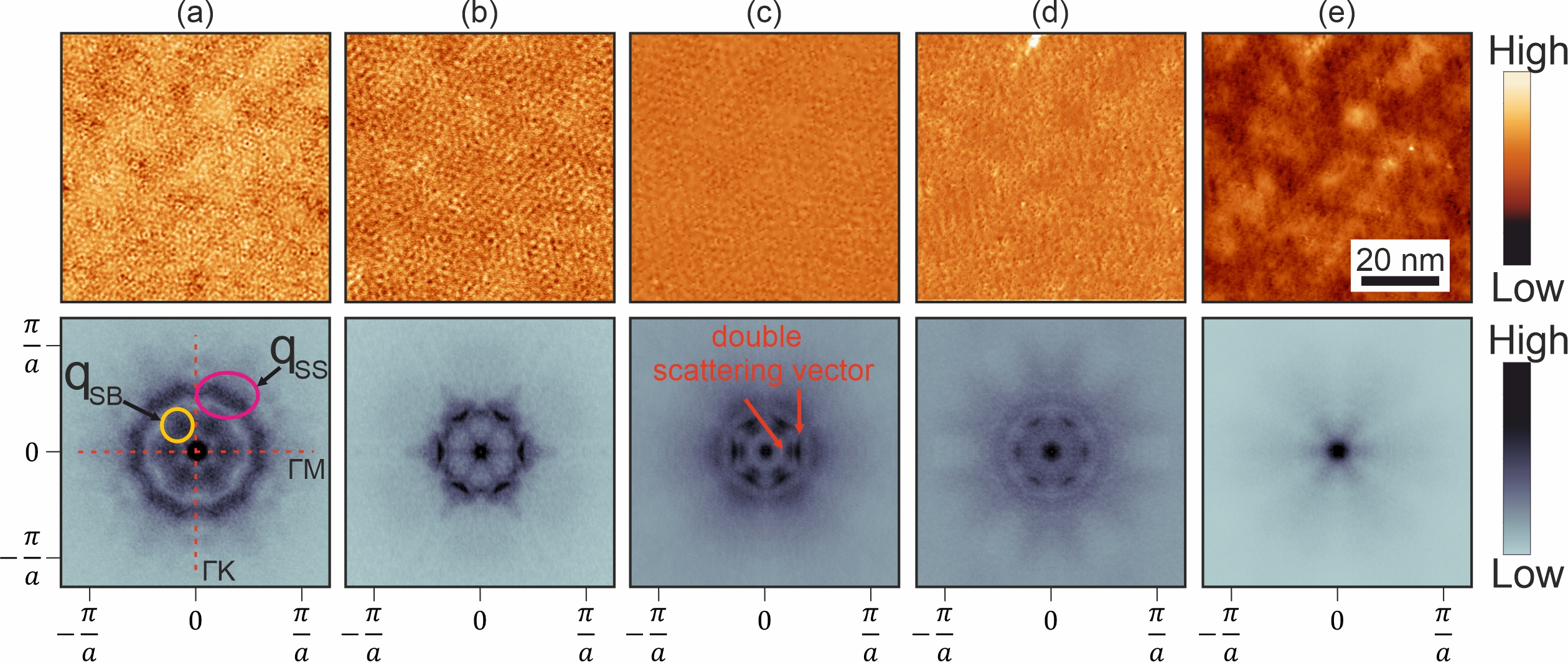}
		\caption{(a)$-$(e) $dI/dU$ maps (top) with corresponding FT-QPI patterns (bottom) at specific energies of 300 mV, 50 mV, $-170$ mV, $-370$ mV, $-470$ mV, respectively, measured on the same surface area. High-symmetry crystallographic directions are labelled in (a). Two distinguishable sets of scattering vectors $q_{SS}$ and $q_{SB}$ are marked by magenta and yellow ellipses, respectively. The in-plane lattice constant value $a$ was derived from topographic data.}
		\label{fig:Fig.5}
	\end{figure*}

	The FT-QPI data obtained at 19 different bias voltages in the range from $300$ mV to $-600$ mV, which is presented in Fig.~\ref{fig:Fig.3}{\color{blue}}, can be classified into 5 representative types according to characteristic observed structures in the FT-QPI patterns as illustrated in Fig.~\ref{fig:Fig.5}.	
	
	To be more precise, the maps shown at high bias voltages from 300 mV to 150 mV as in Fig.~\ref{fig:Fig.5}{\color{blue}{(a)}} exhibit 6 arc-shaped structures of high intensity in the ${\Gamma}$-M direction, characteristic for a Dirac surface state in Bi$_{2}$Te$_{3}$ and labelled as $q_{SS}$ in Fig.~\ref{fig:Fig.5}{\color{blue}{(a)}}. Apart from that, another six-fold symmetric structure of weaker intensity (denoted as $q_{SB}$ in Fig.~\ref{fig:Fig.5}{\color{blue}{(a)}}) collinear with the abovementioned one is clearly visible. At lower energies, in the range [$-70$; 100] mV (Fig.~\ref{fig:Fig.5}{\color{blue}{(b)}}), the FT-QPI patterns only consist of the spots that can be assigned to $q_{SS}$. With decreasing bias voltage the scattering vectors shorten	(see Figs.~\ref{fig:Fig.3} and~\ref{fig:Fig.4}) and become more clearly defined and sharper than those at higher energies. Going further to lower bias voltages from $-120$ mV down to $-320$ mV the scattering vector experiences a splitting into two distinct vectors of nearly the same intensities (see Fig.~\ref{fig:Fig.5}{\color{blue}{(c)}}), while the dispersion becomes weaker, i.e., the peak position becomes almost energy independent (Fig.~\ref{fig:Fig.4}). At the bias voltage of $-370$ mV the scattering vector is again single, not splitted (see Fig.~\ref{fig:Fig.5}{\color{blue}{(d)}}), while at further lower biases the FT-QPI pattern becomes less clear with significantly weaker intensity (see Fig.~\ref{fig:Fig.5}{\color{blue}{(e)}}).
	
	\subsection{QPI analysis }
	\label{subsec:QPI analysis }	
	
\begin{figure}[]
		\centering
		\includegraphics[scale=0.4]{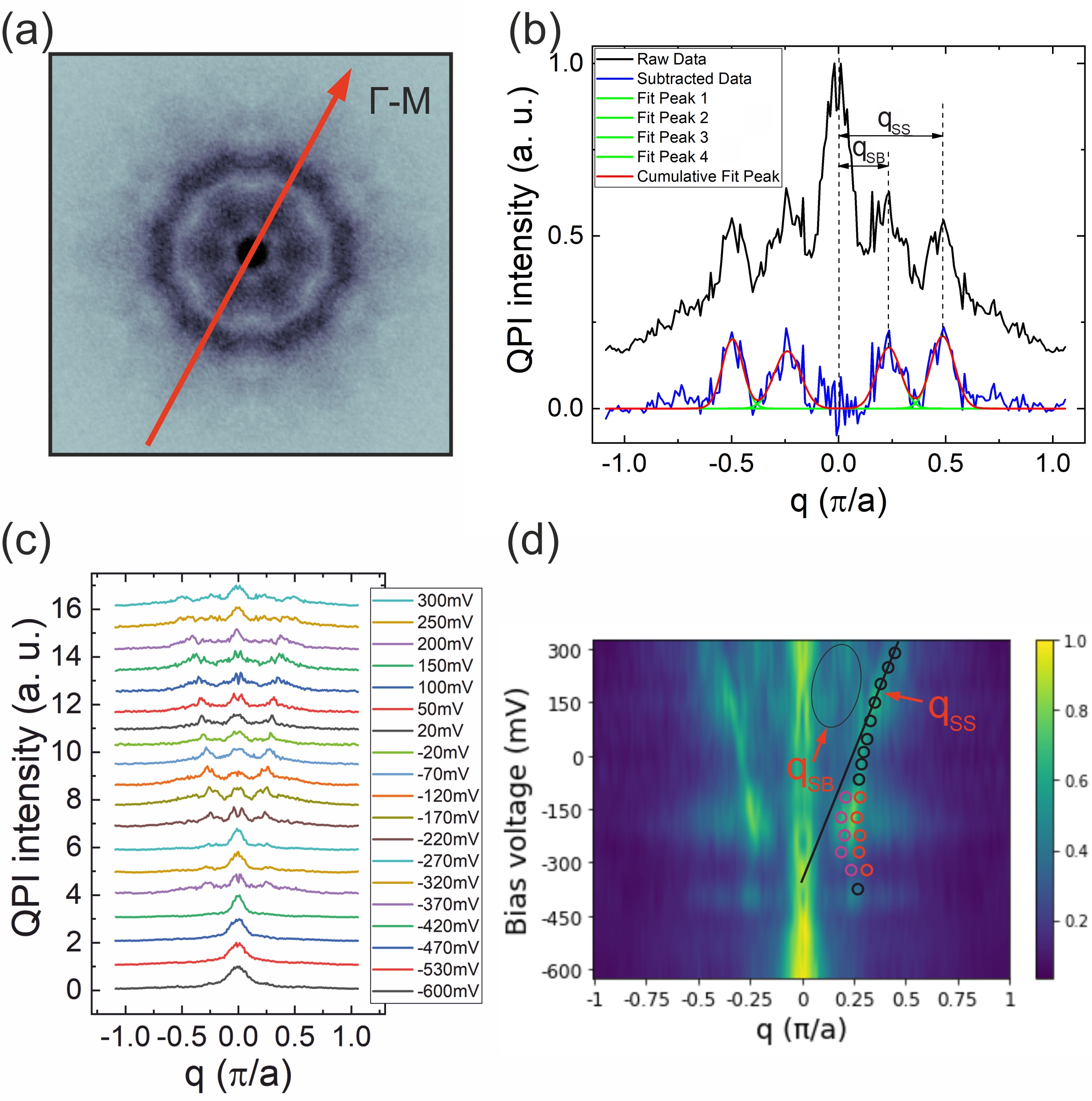}
		\caption{(a) QPI pattern obtained at $U_{bias} = 300$ mV. (b) Intensity profile of the QPI pattern in (a) taken along the direction marked with the red arrow in (a). $q_{SS}$ and $q_{SB}$ are the peak positions corresponding to the outer and the inner scattering vectors in (a). Gaussian fitting of the observed peaks was performed after subtraction of the background with the central peak. (c) QPI intensity profiles along the ${\Gamma}$-M direction at bias voltages from 300 mV down to $-600$ mV. (d) Intensity graph of the QPI energy dispersion obtained from (c). The data extracted from Gaussian fit of the curves in (c) are plotted as circles: black $–$ the outer, dominant vector $q_{SS}$, red and magenta $–$ double scattering vector. The linear fit is indicated by the black line. The black ellipse denotes the inner scattering vector $q_{SB}$.}
		\label{fig:Fig.6}
	\end{figure}

	For the analysis of the FT-QPI data traces along the ${\Gamma}$-M directions were taken for each of the used bias voltages. Gaussian fits of the intensity peaks allowed to extract the lengths of scattering vectors $q_{SS}$ and $q_{SB}$ which are displayed in Fig.~\ref{fig:Fig.6}. 
	
\begin{figure}[]
		\centering
		\includegraphics[scale=0.5]{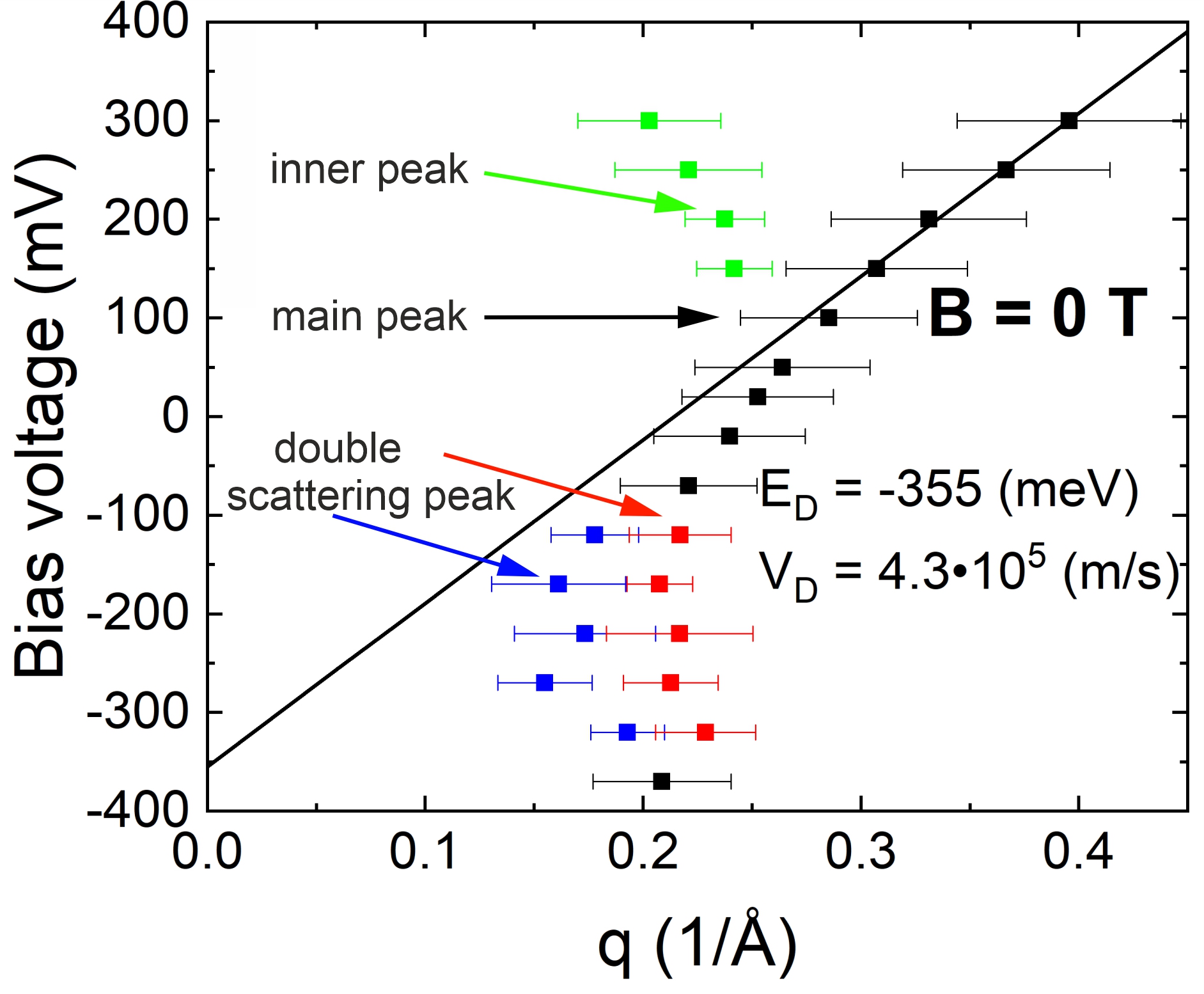}
		\caption{QPI energy dispersion in the ${\Gamma}$-M direction measured at B = 0 T. The data corresponding to the main $\textbf{q}_{SS}$, the inner $\textbf{q}_{SB}$ and the double scattering vectors are plotted in black, green and red/blue colours, respectively. The black line shows the linear fit to the data points in the energy range from 300 meV to 100 meV.}
		\label{fig:Fig.7}
	\end{figure}

	In order to make a linear fit of the energy dispersive QPI scattering vector $q_{SS}$, we took the energy range from 100 mV to 300 mV at which the Dirac velocity derived from the slope of the linear fit of the energy dispersion, as shown in Fig.~\ref{fig:Fig.7}, $v_{D}$ = (4.3 $\pm$ 0.3) $\times$ $10^{5}$ m/s is in a good agreement with	existing data on this compound \cite{netsou2020identifying, PhysRevB.88.161407, doi:10.1126/science.1173034, PhysRevLett.103.266803, PhysRevLett.109.166407}. The extrapolation of the linear fit of $q_{SS}$ at bias voltages from 300 mV to $100$ mV allows to estimate the Dirac point energy to $-355$ meV at which $q = 0$ (see Fig.~\ref{fig:Fig.7}), which was also presumably indicated by the change in the QPI patterns between $-320$ meV and $-370$ meV (see Fig.~\ref{fig:Fig.3}).      
	
	For	further analysis of the scattering vectors in our FT-QPI data we utilize a comparison with ARPES surface band structure data	as well as band structure calculations, in particular, the CEC \cite{PhysRevLett.104.016401, doi:10.1126/science.1173034, PhysRevLett.103.266801, 2009}. 
	
	A simplified first-order approach to model the FT-QPI patterns is provided by joint density of states (JDOS) calculations based on the corresponding band structure data. In this method, the QPI data is simulated by computing the autocorrelation of the DOS in the initial ($\textbf{k}$$\textsubscript{i}$) and final ($\textbf{k}$$\textsubscript{f}$) states (i.e. before and after scattering). In order to take into account the topological spin-momentum locking induced helical in-plane spin texture and the warping induced out-of-plane spin component, we introduce the spin-dependent matrix element:
	\begin{equation}\label{1}
		T(\textbf{q, k}) \propto 1 + \cos({\theta}_{f} - {\theta}_{i}) + s_{z,i}s_{z,f},
	\end{equation}
	where $\theta$$_{f}$ and $\theta$$_{i}$ are the angles that define $\textbf{k}$$\textsubscript{f}$ and $\textbf{k}$$\textsubscript{i}$ in reciprocal space, while $s$$_{z,i}$ and $s$$_{z,f}$ are relative values of the out-of-plane spin component \cite{PhysRevLett.103.266801, Hsieh2009, PhysRevLett.106.216803} for the initial and final states, respectively.
	This allows to obtain spin-dependent scattering probability (SSP) patterns: 
	\begin{equation}\label{2}
		SSP(\textbf{q}) = \int V(\textbf{k})I(\textbf{k})T(\textbf{q, k})V(\textbf{k}+\textbf{q})I(\textbf{k}+\textbf{q})d^2\textbf{k},
	\end{equation}
	which can directly be compared 
	with the experimental FT-QPI patterns. Here, $I(\textbf{k})$ is the momentum dependent DOS, which can e.g., be approximated from ARPES intensity data. In addition, we take into account the three-fold symmetric scattering potential $V$  \cite{Beidenkopf2011, PhysRevB.88.161407, Stolyarov2021} related to the most abundant defect type in our data (see section~\ref{subsec:STM measurements at zero-field }) to obtain the final results of our calculations. Details concerning the effect of the implementation of the spin texture and the scattering potential are presented in Figs.~\ref{fig:Fig.16} and~\ref{fig:Fig.17}.
	
\begin{figure}[H]
		\centering
		\includegraphics[scale=2.2]{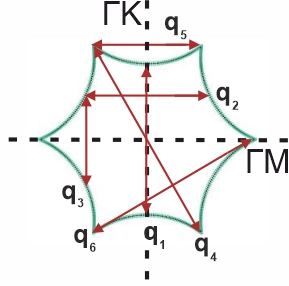}
		\caption{Scattering geometry for the hexagonally warped surface state CEC in Bi$_{2}$Te$_{3}$. The main scattering vectors \textbf{q}$_{1}$$-$\textbf{q}$_{6}$ connect points of the extremal curvature on the CEC. }
		\label{fig:Fig.8}
	\end{figure}

	In order to relate the implications of the FT-QPI patterns to the band structure, qualitative schemes of the CECs for Bi$_{2}$Te$_{3}$ were derived according to the ARPES data \cite{PhysRevLett.104.016401, doi:10.1126/science.1173034} serving as a starting point for the analysis. The schematic illustration of the scattering geometry for the warped hexagonal shape of the CEC in Bi$_{2}$Te$_{3}$ is shown in Fig.~\ref{fig:Fig.8}. For the distinct discussion we consider 6 exemplary scattering vectors resulting from the geometry of the CEC, according to \cite{PhysRevB.80.245439}.  
	
	In order to implement the SSP calculations, the DOS variation on each of the CECs was qualitatively modeled with a Gaussian distribution, considering the DOS distribution experimentally determined by ARPES. Thereby the width of the gaussians was adjusted as to obtain reasonable simulation results (see more details in Fig.~\ref{fig:Fig.18}). 
	
	\begin{figure*}[]
		\centering
		\includegraphics[width=0.95\textwidth]{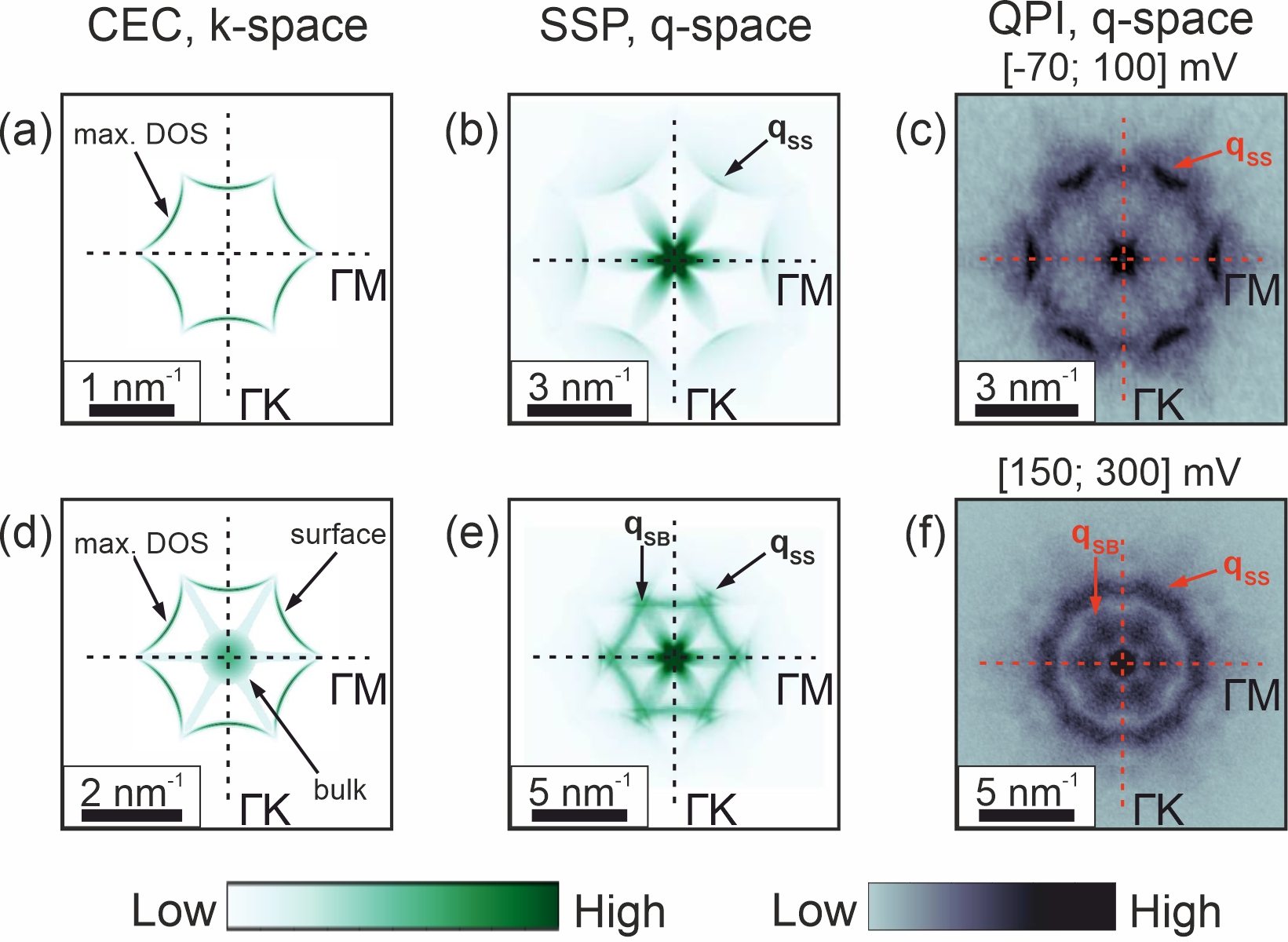}
		\caption{(a)$,$(d) CECs considered for different energy ranges according to the ARPES data \cite{PhysRevLett.104.016401, doi:10.1126/science.1173034}. (b)$,$(e) SSP calculations corresponding to the CECs in (a)$,$(d), respectively. (c)$,$(f) QPI patterns measured in our work with the energy ranges indicated on top of each figure. The high-symmetry directions and the scalebars are marked for all the presented figures. The left and the right colorbars correspond to the data in the two columns on the left side and the right column, respectively.}
		\label{fig:Fig.9}
	\end{figure*}
	
	We start our comparison of the experimental data with the SSP calculations for the hexagonally warped surface state CEC with no bulk contribution, as shown in Fig.~\ref{fig:Fig.9}{\color{blue}{(a)}}. The maximal DOS is on the warped hexagon, in the ${\Gamma}$-K direction, according to the ARPES data \cite{PhysRevLett.104.016401, doi:10.1126/science.1173034}. This leads to the six-fold symmetric SSP pattern in Fig.~\ref{fig:Fig.9}{\color{blue}{(b)}}, where the outer structure arising from backscattering with the vectors	\textbf{q}$_{1}$, \textbf{q}$_{4}$ is strongly suppressed	(see more details in Fig.~\ref{fig:Fig.16}). The arc-shaped six-fold symmetric structures of high intensity in the ${\Gamma}$-M direction are in a reasonable agreement with the QPI data measured at $U_{bias}$ = [$-70; 100$] mV shown in Fig.~\ref{fig:Fig.9}{\color{blue}{(c)}}, thus, they are associated with \textbf{q}$_{5}$ or \textbf{q}$_{2}$ scattering processes forming the aforementioned $\textbf{q}_{SS}$. It is, however, difficult to disentangle which of these two vectors is more relevant. It is important to note that our QPI simulation result exhibits a flower-like shaped pattern in the ${\Gamma}$-M direction at scattering vectors shorter than $\textbf{q}_{5}$ (see Fig.~\ref{fig:Fig.9}{\color{blue}{(b)}}). The origin of this pattern comes from scattering within a single arc on the hexagonally warped CEC, thus, it cannot be totally suppressed. However, by	calculating the SSP with optimized resolution in the \textbf{q}-space, leading to the intensity suppression at its lobes and intensity enhancement at its center, the flower-like pattern can become consistent with our experimental data in Fig.~\ref{fig:Fig.9}{\color{blue}{(c)}}.
	
	At higher energies, from 150 meV to 300 meV, the surface state CEC alone does not suffice to describe the experimentally acquired QPI patterns at this energy range (see more details in Fig.~\ref{fig:Fig.19}). Therefore, we take not only the hexagonal warping of the surface state CEC explicitly into account but also approximate the bulk conduction band as a continuum centered at the ${\Gamma}$-point with the DOS smoothly decaying away from ${\Gamma}$ and extending in the ${\Gamma}$-M direction, as shown in Fig.~\ref{fig:Fig.9}{\color{blue}{(d)}}, mimicking the band structure data \cite{doi:10.1126/science.1173034}. According to the experimental data in Fig.~\ref{fig:Fig.9}{\color{blue}{(f)}}, the arc-shaped structures \textbf{q}$_{SS}$ appear slightly less well defined than those at lower energies which is resembled by the SSP simulation involving bulk states (see Fig.~\ref{fig:Fig.9}{\color{blue}{(e)}}). 
	Apart from that, in the considered energy range the QPI shown in Fig.~\ref{fig:Fig.9}{\color{blue}{(f)}} exhibits an enhanced non-dispersive intensity in the ${\Gamma}$-M direction at scattering vectors shorter than \textbf{q}$_{SS}$ (the inner ring-like structure). Thus, the origin of these peaks cannot be explained in terms of the surface state contributions to the CEC alone, and their existence results from bulk contributions to the scattering. In particular, this inner ring-like structure $\textbf{q}_{SB}$ arises from the bulk-to-surface scattering in the bulk conduction band, as shown in Fig.~\ref{fig:Fig.19}. Note that this surface-bulk scattering scenario was previously proposed to interpret the experimental QPI data of Bi$_{2}$Se$_{3}$ \cite{PhysRevLett.107.056803}. 
	
	In contrast to the energy dispersion of the Dirac surface states, the dispersion of the QPI features in our experimental data deviates from the linear one below $50$ meV and cannot be observed below $-70$ meV (see Fig.~\ref{fig:Fig.7}). In light of the inconsistency between the linearly dispersing surface state bands and the experimentally observed lack of dispersion in the QPI data, we propose to involve bulk, in particular, helical so-called bulk-surface hybridized \cite{Hedayat2021, Hsu2014, PhysRevB.88.161407} states in scattering processes for energies below $-70$ meV, too. Within this scenario, the two observed non-dispersive scattering modes (above $100$ mV and below $-70$ mV) allow us to roughly estimate the bulk energy gap range in the measured samples as [$-70$; 100] $\pm$ 25 mV, and the gap size is consistent with the literature data on Bi$_{2}$Te$_{3}$ \cite{doi:10.1126/science.1173034}. 
	
\begin{figure*}[]
		\centering
		\includegraphics[width=0.98\textwidth]{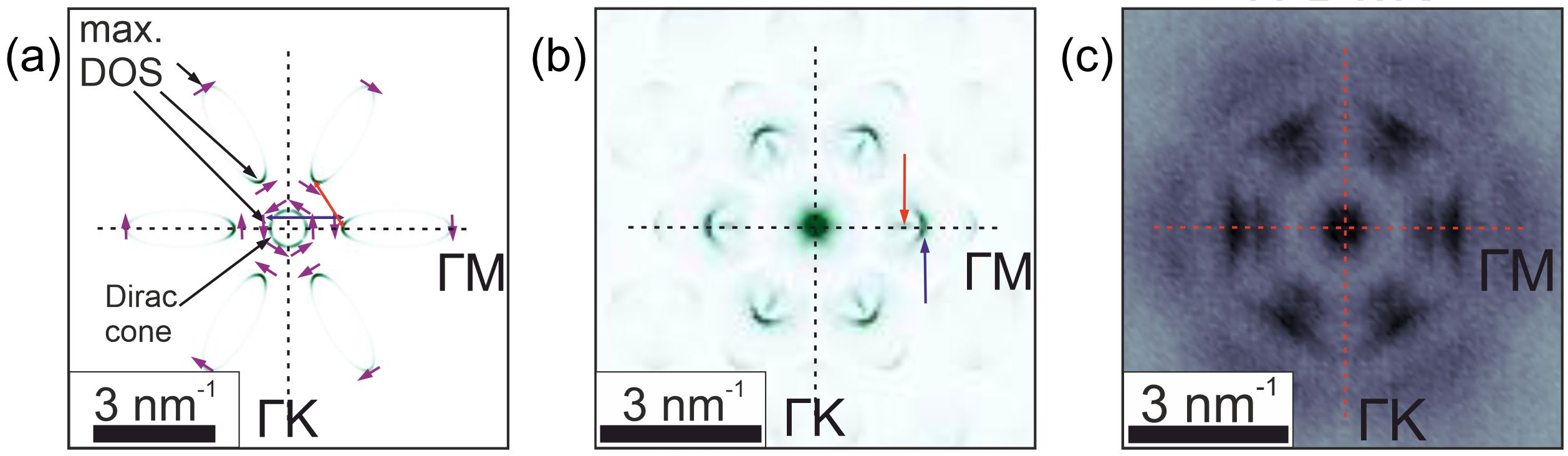}
		\caption{(a) CEC for the Dirac surface state shown as the circle and the surface state band merging with the bulk valence band as six-fold symmetric elliptically shaped pockets. All the states are spin-polarized, the spin texture is represented by magenta arrows. Maximal DOS is set in the ${\Gamma}$-M direction for all the parts of the CEC. Two dominant scattering vectors are indicated by red and blue arrows. (b) SSP pattern for the CEC plotted in (a). Two scattering peaks coming from scattering processes shown in (a) are denoted by red and blue arrows. (c) Exemplary QPI pattern exhibiting the double structure measured at $U_{bias} = -170$ mV.}
		\label{fig:Fig.10}
	\end{figure*}

	As mentioned before, our QPI data measured at \\$U_{bias}$ = [$-320$; $-120$] mV (see Fig.~\ref{fig:Fig.3}) exhibit double intensity peaks in the ${\Gamma}$-M direction at $q\approx0.16$ 1/\AA{} and 0.22 1/\AA{}. In order to interpret this	experimental finding, we model the surface CEC in the bulk valence band energy regime and the DOS distribution according to the known surface band structure of Bi$_{2}$Te$_{3}$ \cite{doi:10.1126/science.1173034, PhysRevLett.104.016401, PhysRevB.88.161407, Stolyarov2021} (see Fig.~\ref{fig:Fig.10}{\color{blue}{(a)}}). Apart from the non-warped Dirac cone, the CEC has six-fold symmetrical pockets of elliptical shape extending in the ${\Gamma}$-M direction and representing the bulk-surface hybridized states \cite{Hedayat2021, Hsu2014, PhysRevB.88.161407}. The maximal DOS is set in the ${\Gamma}$-M direction for all the parts of the CEC, and the spin texture is modeled according to \cite{Hsieh2009}. Upon calculating the SSP, we obtain two scattering peaks in the ${\Gamma}$-M direction consistent with our experimental data in the corresponding energy range. Fig.~\ref{fig:Fig.10}{\color{blue}{(b)}} illustrates the corresponding SSP pattern with two six-fold symmetric structures in the ${\Gamma}$-M direction, resembling our experimental QPI patterns (see Fig.~\ref{fig:Fig.10}{\color{blue}{(c)}}). However, there is a strong difference in the shape of the abovementioned six-fold symmetric structures (see Figs.~\ref{fig:Fig.10}{\color{blue}{(b)}} and~\ref{fig:Fig.10}{\color{blue}{(c)}}), and, in the model, they exhibit an energy dispersion, in contrast to our experimental data. Therefore, either a different dispersion of the bulk-surface hybridized states as described in Ref. 22 or another, unknown, mechanism behind the non-dispersive modes should be considered \cite{Footnote}.
	
	At yet lower energies ($U_{bias}$ = $-370$ mV), the double QPI structure disappears, and a single scattering vector is detected in the ${\Gamma}$-M direction (see Fig.~\ref{fig:Fig.5}{\color{blue}{(a)}}). In the above scenario which involves the bulk-surface hybridized states, this indicates an accidental match of the length of the involved scattering vectors. 
	
	Finally, the observed very faint experimental QPI patterns for $U_{bias}$ = [$-600$; $-420$] mV showing a star-like shape in the ${\Gamma}$-M direction can be interpreted by scattering processes within the bulk valence band at the energies significantly below the Dirac point. 
	
	Comparing the energy scales from the QPI data in our work with the former QPI and ARPES studies in Bi$_{2}$Te$_{3}$ \cite{netsou2020identifying, PhysRevB.88.161407, PhysRevLett.104.016401, Stolyarov2021, doi:10.1126/science.1173034, PhysRevLett.103.266803, PhysRevLett.109.166407}, it is clear that there is a difference in the bulk energy gap location with respect to the Dirac point which are separated by around 250 meV (counting distance between the Dirac point and the top of the bulk valence band) in our data unlike 100$-$200 meV in the literature. The reason for this discrepancy is unknown. There might be a relation with the relatively high density of impurities in our samples. These impurities could give rise to an effective doping and band structure modifications of the compound. Note that the modified Dirac cone dispersion with respect to the bulk states naturally implies different bulk-surface hybridized states as commonly considered \cite{PhysRevB.88.161407} (see discussion of the non-dispersive splitted QPI peaks above).

	\subsection{Magnetic field measurements}	
	\label{subsec:Magnetic field measurements }		

	In order to scrutinize the influence of an external magnetic field on the topological properties manifested by the suppressed backscattering we performed measurements in magnetic fields up to 15 T. This allows to compare the QPI features found in the zero-field data, which were discussed in sections~\ref{subsec:STM measurements at zero-field } and~\ref{subsec:QPI analysis }, to the in-field QPI features caused by the presence of the magnetic field, which, if strong enough, could break the time-reversal symmetry and the topological protection of the surface states. 
	
\begin{figure}[]
		\centering
		\includegraphics[scale=0.85]{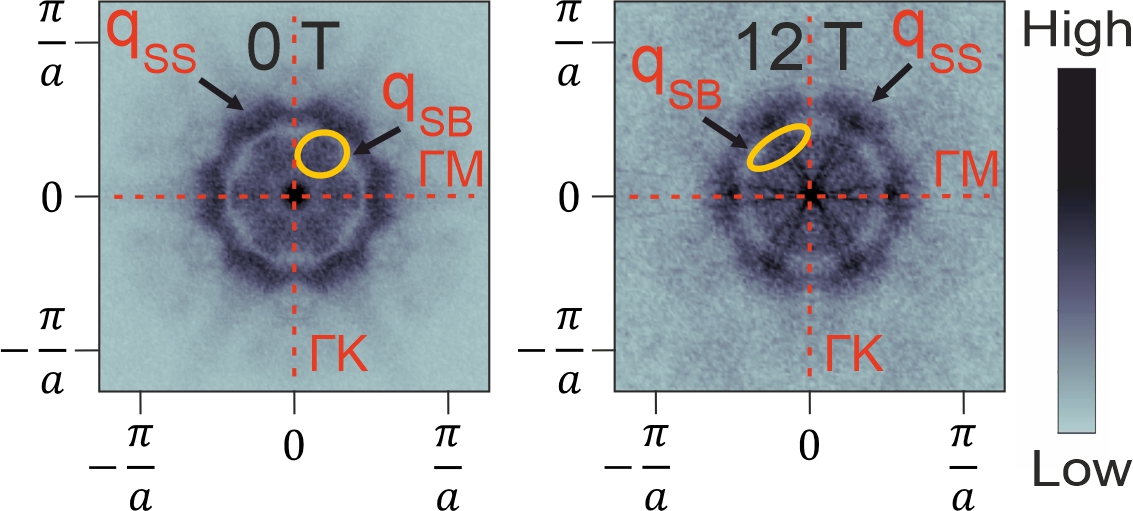}
		\caption{Representative FT-QPI patterns measured at \\$U_{bias}$ = 250 mV at $B$ = 0 T and 12 T. The outer intensity spots correspond to the vector \textbf{q}$_{SS}$ for each of the patterns. The inner intensity spot, which is visible only at 300 mV, 250 mV and 200 mV, corresponds to the scattering vector \textbf{q}$_{SB}$ and is indicated by the yellow ellipse.}
		\label{fig:Fig.11}
	\end{figure}

	All	the $dI/dU$ maps in the magnetic field of 12 T were acquired on the same 60 nm $\times$ 60 nm area at 9 bias voltages from 300 mV to $-50$ mV (see Fig.~\ref{fig:Fig.20}). All the FT-QPI patterns measured at $B = 12$ T are presented in Fig.~\ref{fig:Fig.21} in comparison with the corresponding data obtained at $B = 0$ T. A representative FT-QPI pattern measured at $U_{bias}$ = 250 mV is depicted in Fig.~\ref{fig:Fig.11} in comparison with that obtained at $B = 0$ T. One can notice slight differences in the FT-QPI intensity distributions which could be caused by different combinations of samples and tips in different zero-field and in-field experiments. As found in the QPI data measured at $B = 0$ T, the dominant six-fold symmetric scattering vector at $B = 12$ T pointing in the ${\Gamma}$-M direction exhibits a continuous shortening upon decreasing the bias voltage (see the zero-field data above). This scattering vector exists at all the data measured with the bias voltage varying from 300 mV to $-50$ mV at $B = 12$ T. In addition, the bulk scattering mode observed at high energies in the zero-field data has been detected at $B = 12$ T, too (see section~\ref{subsec:QPI analysis }).
	
\begin{figure}[]
		\centering
		\includegraphics[scale=0.42]{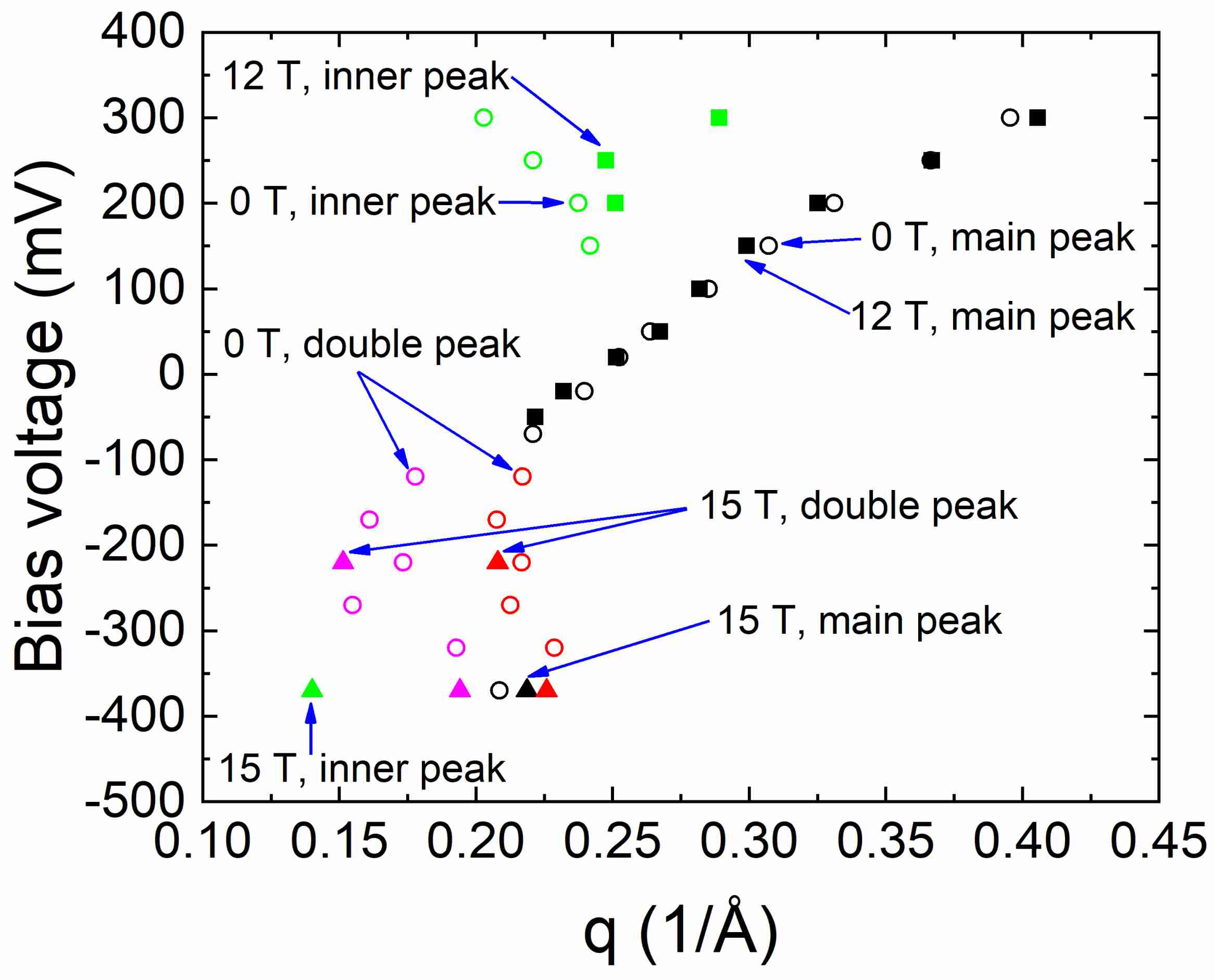}
		\caption{QPI energy dispersion measured at $B = 0$ T, 12 T and 15 T. Circles, squares and triangles represent the data measured in the zero field, $B = 12$ T and $B = 15$ T, respectively. The data corresponding to the main $\textbf{q}_{SS}$, the inner $\textbf{q}_{SB}$ and the splitted scattering vectors are plotted in black, green and red/magenta colours, respectively.}
		\label{fig:Fig.12}
	\end{figure}

	The linear fit of the energy dispersion of the dominant scattering vector (see Fig.~\ref{fig:Fig.12}) gives the estimation of the Dirac point situated at $U_{bias}$ = $-384$ $\pm$ $59$ mV with the Dirac velocity $v_{D}$ = ($4.5$ $\pm$ $0.3) \times 10^{5}$ m/s which are both in agreement with the corresponding values at $B = 0$ T being within the uncertainty of the experiment. No backscattering peaks were observed in the QPI data which may reinforce the assumption that even the external field of $B = 12$ T was not strong enough to break the time-reversal symmetry. Overall, no significant changes in the FT-QPI patterns at $B = 12$ T with respect to that at $B = 0$ T were found as can be inferred from Fig.~\ref{fig:Fig.21}. This can be explained by the fact that the Zeeman interaction even at $B = 15$ T is around 0.87 meV which is significantly weaker than the spin-orbit interaction in Bi$_{2}$Te$_{3}$ (the spin-orbit coupling constant $\lambda$ = 1.25 eV for Bi and 0.49 eV for Te, respectively \cite{Wittel1974, Zhang2009}). Thus, only a very small effect $\sim$0.001 of the external magnetic field could be expected.
	
	The $dI/dU$ maps were also measured at the maximal achievable field of 15 T at the bias voltages of $-220$ mV and $-370$ mV. The FT-QPI patterns are presented in Fig.~\ref{fig:Fig.22} together with the corresponding data measured at $B = 0$ T. However, even with a better momentum resolution provided by 200 nm $\times$ 200 nm surface areas which were different in each of the presented data on the same sample, no substantial changes with respect to the zero-field data at those energies were found. The energy dispersion $E(q)$ obtained at $B = 15$ T is also shown in Fig.~\ref{fig:Fig.12} along with the QPI dispersion measured at $B = 12$ T and $B = 0$ T.	
	
	\section{Conclusions}
	
	In summary, we have carried out a comprehensive QPI study of disordered Bi$_{2}$Te$_{3}$ crystals in a relatively large energy range [$-600$; 300] meV and in magnetic field. We have reported the energy dependence of the FT-QPI patterns which have been analysed through spin-dependent JDOS calculations providing a reasonable agreement with the experiment to some extent. The linear energy-momentum dispersion and the absence of the backscattering were clearly demonstrated proving the helical spin texture of the material. The QPI data allowed us to roughly compare the electronic band structure of the material with its known band structure based on our experimental data. In particular, we interpreted the observed non-dispersive scattering vectors as originating from the involvement of the bulk states with the possible indication of the bulk energy gap size of 170 $\pm$ 25 meV consistent with its known value, which can serve as a basis for future investigations. We have also measured the QPI data in the magnetic field from which we conclude that the topological protection of the surface state survives even at $B = 12$ T due to the strong spin-orbit coupling in Bi$_{2}$Te$_{3}$.
	
	\begin{acknowledgments}		
		
		This work has been supported by the European Research Council (ERC) under the European Union’s Horizon 2020 research and innovation programme (Grant Agreement No. 647276-MARS-ERC-2014-CoG) and by the Deutsche Forschungsgemeinschaft (DFG, German Research Foundation) through SFB 1143 (project-id 247310070).
		
	\end{acknowledgments}
	
	\appendix
	\numberwithin{equation}{section}
	
	\section*{APPENDIX A: ADDITIONAL EXPERIMENTAL DATA ON Bi$_{2}$Te$_{3}$ }
	\label{A}
	\renewcommand{\theequation}{A\arabic{equation}}
	
\begin{figure}[H]
		\centering
		\includegraphics[scale=0.75]{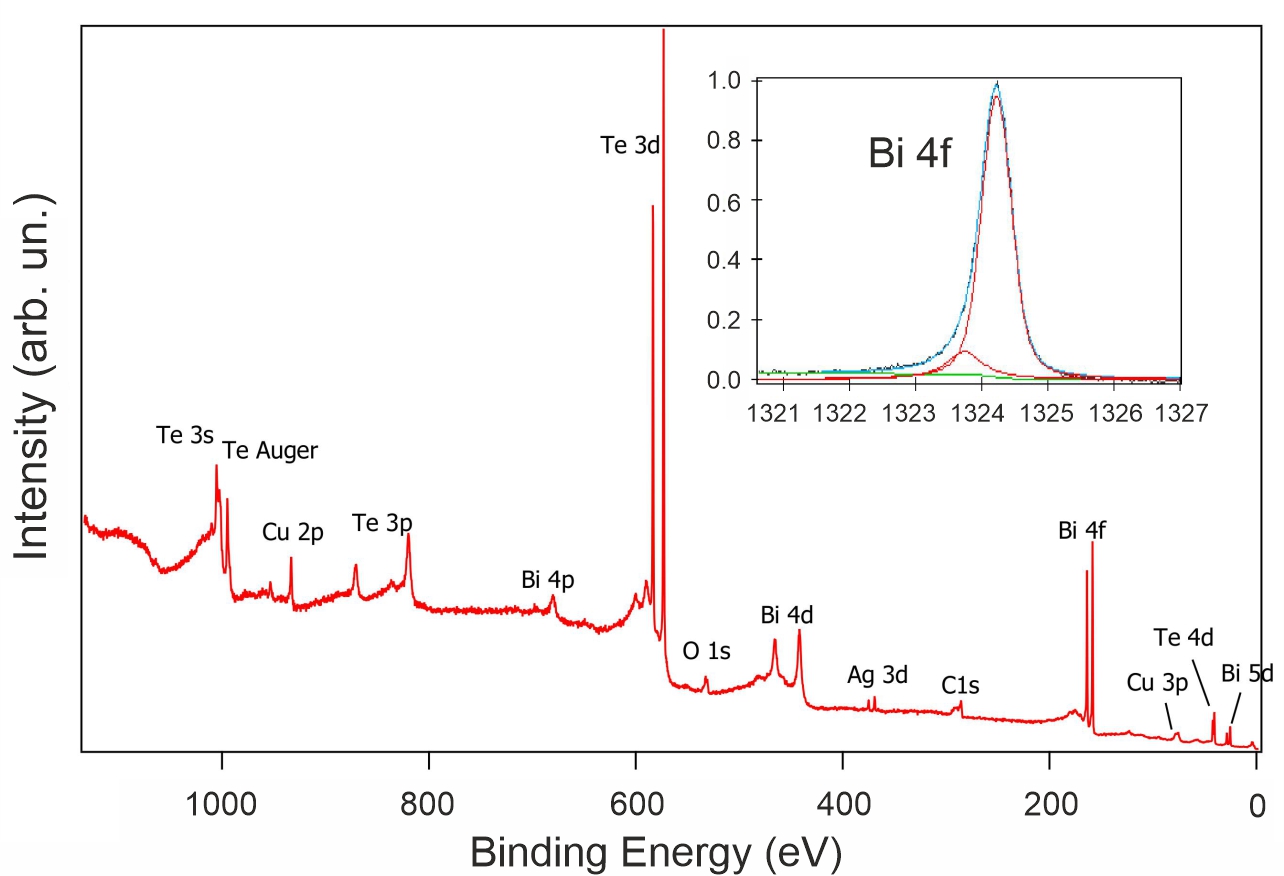}
		\caption{XPS spectra of Bi$_{2}$Te$_{3}$ measured in our work. The inset shows two components of the Bi $4f$ peak (red curves) derived from the fitting of the experimental data (black curve). The cyan curve is the sum of the two red curves. Note that the energy scale in the inset displays the kinetic energy of electrons and is different from the one shown in the main spectrum.}
		\label{fig:Fig.13}
	\end{figure}

	X-ray photoelectron spectroscopy (XPS) measurements have been performed to exclude possible extrinsic doping of the crystals. The XPS analysis reveals the absence of extrinsic dopants (which confirms the intrinsic origin of the defects) and no deviation from the stoichiometry of this compound. Note that there are relatively small contents of oxygen and carbon contaminants (see Fig.~\ref{fig:Fig.13}{\color{blue}}) which are often observed in XPS experiments, also copper and silver peaks in the spectrum are related to the sample holder and the component of the conducting glue, respectively. At first glance, no substantial differences were recognized in the obtained spectrum compared to the previously measured XPS spectra on Bi$_{2}$Te$_{3}$ \cite{Shallenberger2019}. However, a closer look at the relatively narrow Bi $4f$ peak reveals slight deviations from its ideal shape. In order to fit the photoemission spectrum by Voigt profiles, a second component of the Bi $4f$ peak needs to be taken into account, as shown in Fig.~\ref{fig:Fig.13}. This suggests modified chemical environments of Bi atoms, which can be evidenced by the disorder in the STM topographic data (see the main text). Meanwhile, the additional contribution to the Bi $4f$ peak was also detected by XPS in \cite{Concepcion2018} where it was caused by Bi surface oxides, the latter could occur in our crystals as well due to their exposure to air prior to the STM measurements.
	
\begin{figure}[H]
		\centering
		\includegraphics[scale=0.43]{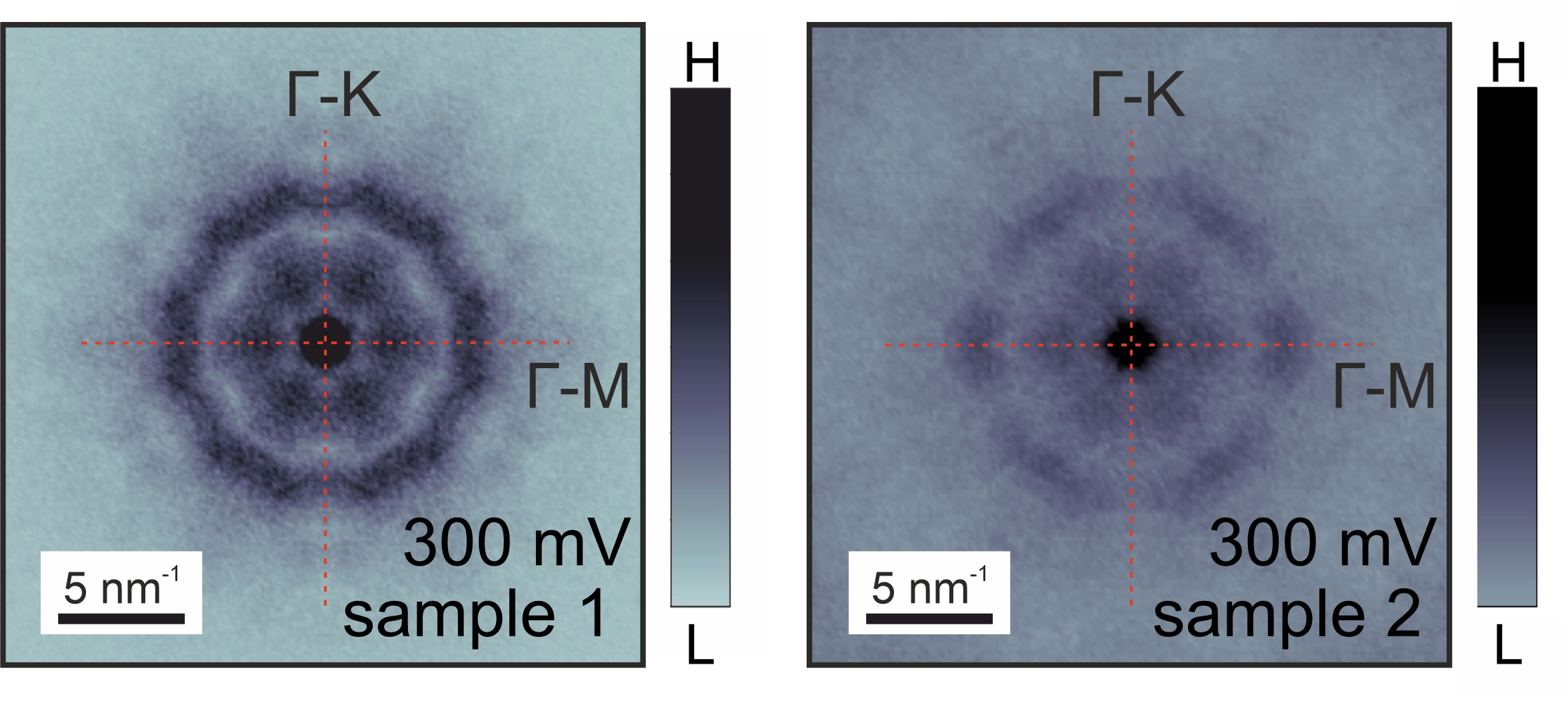}
		\caption{Representative symmetrized FT-QPI patterns measured on different samples at the same $U_{bias} = 300$ mV. The data obtained on the "sample 1" are presented in the main text. The high-symmetry directions are indicated with red lines.}
		\label{fig:Fig.14}
	\end{figure}

	We have obtained QPI data on 5 different samples, and the corresponding FT-QPI patterns are qualitatively similar at respective energies. One example is presented in Fig.~\ref{fig:Fig.14}, where the six-fold symmetric QPI structures discussed in the main text exist on 2 different samples, but there are slight differences in the FT-QPI intensity distributions, probably caused by different combinations of samples and STM tips.
	
\begin{figure}[H]
		\centering
		\includegraphics[scale=0.4]{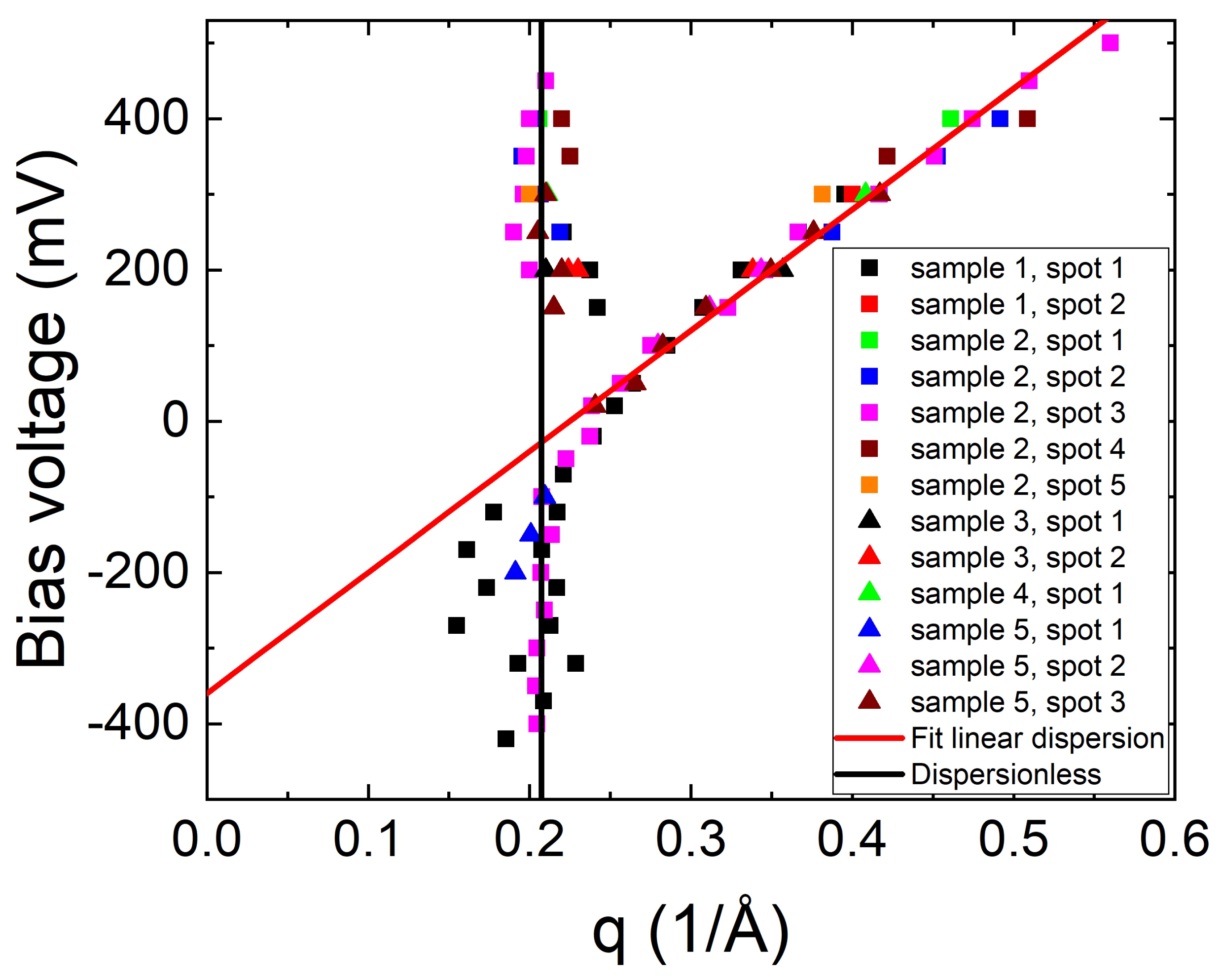}
		\caption{QPI energy dispersion in the ${\Gamma}$-M direction obtained at $B = 0$ T on different samples and different surface areas on the same samples.}
		\label{fig:Fig.15}
	\end{figure}

	The lengths of the dominant scattering vectors extracted from the QPI data on 5 measured samples and, in addition, on different surface areas of the same samples are plotted in Fig.~\ref{fig:Fig.15}. From that one can clearly argue that the data are sample independent within the batch under investigation. The linear fit for the dominant scattering vector $\textbf{q}_{SS}$ for 5 samples is marked by the red line resulting in the Dirac point energy of $-360$ $\pm$ 40 mV and $v_{D}$ = (4.3 $\pm$ 0.4) $\times$ $10^{5}$ m/s. The non-dispersive scattering mode with the average $q$ $\approx$ 0.21 1/\AA~is indicated by the vertical black line. 
	
	\section*{APPENDIX B: DETAILS OF THE QPI SIMULATIONS }
	\label{B}
	\renewcommand{\theequation}{B1}
	
\begin{figure}[H]
		\centering
		\includegraphics[scale=0.9]{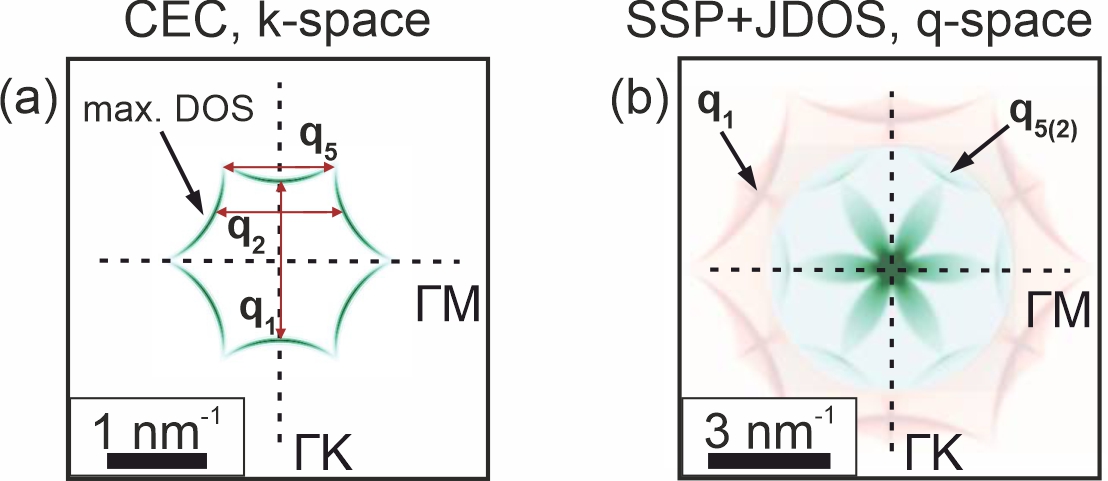}
		\caption{(a) The hexagonally warped surface state CEC of Bi$_{2}$Te$_{3}$. (b) The corresponding SSP pattern (green color), on which the long scattering vectors of the JDOS pattern (red color) are overlayed. The dominant scattering vectors for the SSP (\textbf{q}$_{5(2)}$) and for the outer structure of the JDOS (\textbf{q}$_{1}$ $-$ backscattering) are marked on both figures.}
		\label{fig:Fig.16}
	\end{figure}

	As mentioned in the main text, the SSP pattern for the hexagonally warped surface CEC consists of the six-fold symmetric scattering vector $\textbf{q}_{SS}$ which is a combination of the scattering vectors $\textbf{q}_{2}$ and $\textbf{q}_{5}$. Backscattering processes, mainly corresponding to the scattering vector $\textbf{q}_{1}$ for the CEC plotted in Fig.~\ref{fig:Fig.16}{\color{blue}{(a)}}, are suppressed due to the spin-momentum locking. They can be clearly revealed by calculating the JDOS pattern which is displayed in red color in Fig.~\ref{fig:Fig.16}{\color{blue}{(b)}}. 
	
	We used the three-fold symmetric scattering potential in the SSP calculations, which comes from the dominant defect type in our STM data, in the QPI simulations. We model it as follows: 
	\begin{equation}\label{1}
		\resizebox{0.42\textwidth}{!}{$V = const\times
			\begin{cases}
				exp(-\dfrac{\abs{\theta - \pi/6}}{\pi/48}),
				& \text{ $\theta=[-\dfrac{\pi}{6}; \dfrac{\pi}{2}]$},\\
				exp(-\dfrac{\abs{\theta - 5\pi/6}}{\pi/48}),
				& \text{ $\theta=[\dfrac{\pi}{2}; \dfrac{7\pi}{6}]$},\\
				exp(-\dfrac{\abs{\theta - 3\pi/2}}{\pi/48}),
				& \text{ $\theta=[\dfrac{7\pi}{6}; \dfrac{11\pi}{6}]$},\\
			\end{cases}
			$}
	\end{equation}  
	where the angle $\theta$ defines the vector \textbf{k} in reciprocal space and was introduced in the main text. 
	
\begin{figure}[H]
		\centering
		\includegraphics[scale=0.33]{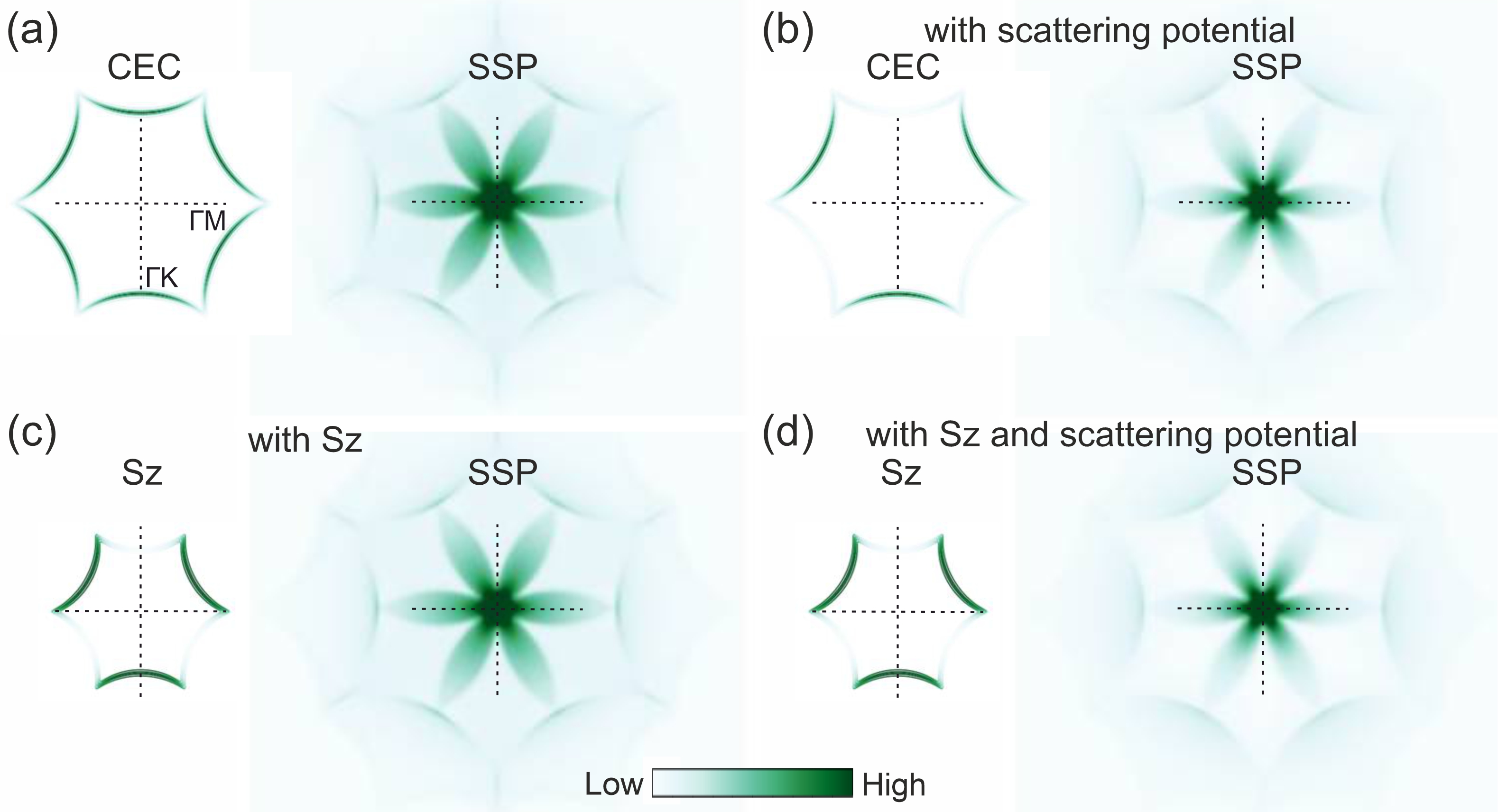}
		\caption{Schematic illustration of the SSP calculations with the out-of-plane spin component and the scattering potential. The high-symmetry directions and the scalebars are the same for all the presented figures.}
		\label{fig:Fig.17}
	\end{figure}

	Fig.~\ref{fig:Fig.17} represents an example of the effect of the influence of the scattering potential and the out-of-plane spin component on the SSP pattern for the hexagonally warped CEC. Fig.~\ref{fig:Fig.17}{\color{blue}{(a)}} shows the result, when both are not taken into account. Fig.~\ref{fig:Fig.17}{\color{blue}{(b)}} corresponds to the case, when only the scattering potential is considered, and results in the suppressed SSP intensity of the inner flower-like shaped pattern. Fig.~\ref{fig:Fig.17}{\color{blue}{(c)}} illustrates a weaker suppression of this inner pattern which is caused by including the out-of-plane spin component. The final result is plotted in Fig.~\ref{fig:Fig.17}{\color{blue}{(d)}} and was obtained taking into account both the scattering potential and the out-of-plane spin component, and the inner flower-like pattern becomes significantly suppressed compared to Fig.~\ref{fig:Fig.17}{\color{blue}{(a)}}.       
	
\begin{figure}[H]
		\centering
		\includegraphics[scale=0.27]{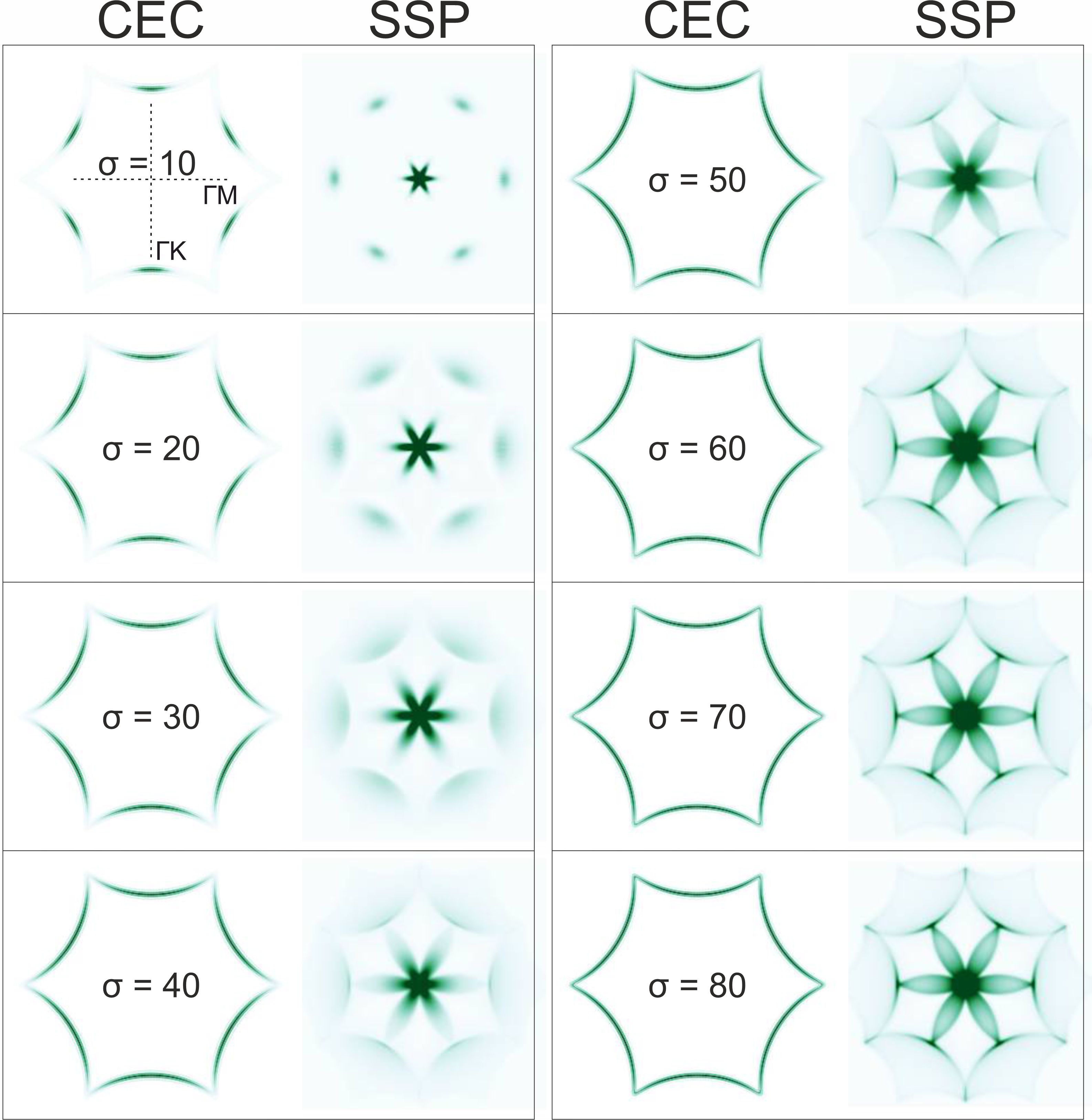}
		\caption{Hexagonally warped CECs with different widths $\sigma$ of the Gaussian distribution of the DOS and the corresponding SSP patterns. The high-symmetry directions and the scales are the same for all the presented figures. The color scale is the same as that in Fig.~\ref{fig:Fig.17}.}
		\label{fig:Fig.18}
	\end{figure}

	We used the Gaussian distribution of the DOS on each of the 6 arcs of the CEC, e.g., for $\theta=[0; \pi/3]$ it is proportional to $exp(-\dfrac{(i-n/12)^{2}}{2\sigma^{2}})$, where $i$ and $n = 1000$ are the index of a certain point and the total number of points on the modeled CEC. Fig.~\ref{fig:Fig.18} shows the evolution of the SSP patterns for the hexagonally warped surface CEC when the width $\sigma$ of the Gaussian distribution of the DOS is changing from 10 to 80. We found the best match to our QPI data at $\sigma$ = [30; 50]. The simulation results presented in the main text were obtained with \\$\sigma$ = 40, which corresponds to the DOS ratio of about 10:1 between the center of the arc of the CEC and the corner. In particular, at lower $\sigma$ values the shape of the SSP scattering peaks differs from that obtained in the experiment, and at higher values of $\sigma$ the inner flower-like shaped pattern becomes more intense (because it results from scattering within a single arc, which has a larger DOS at the corners at higher $\sigma$ values).    
	
\begin{figure}[H]
		\centering
		\includegraphics[scale=0.59]{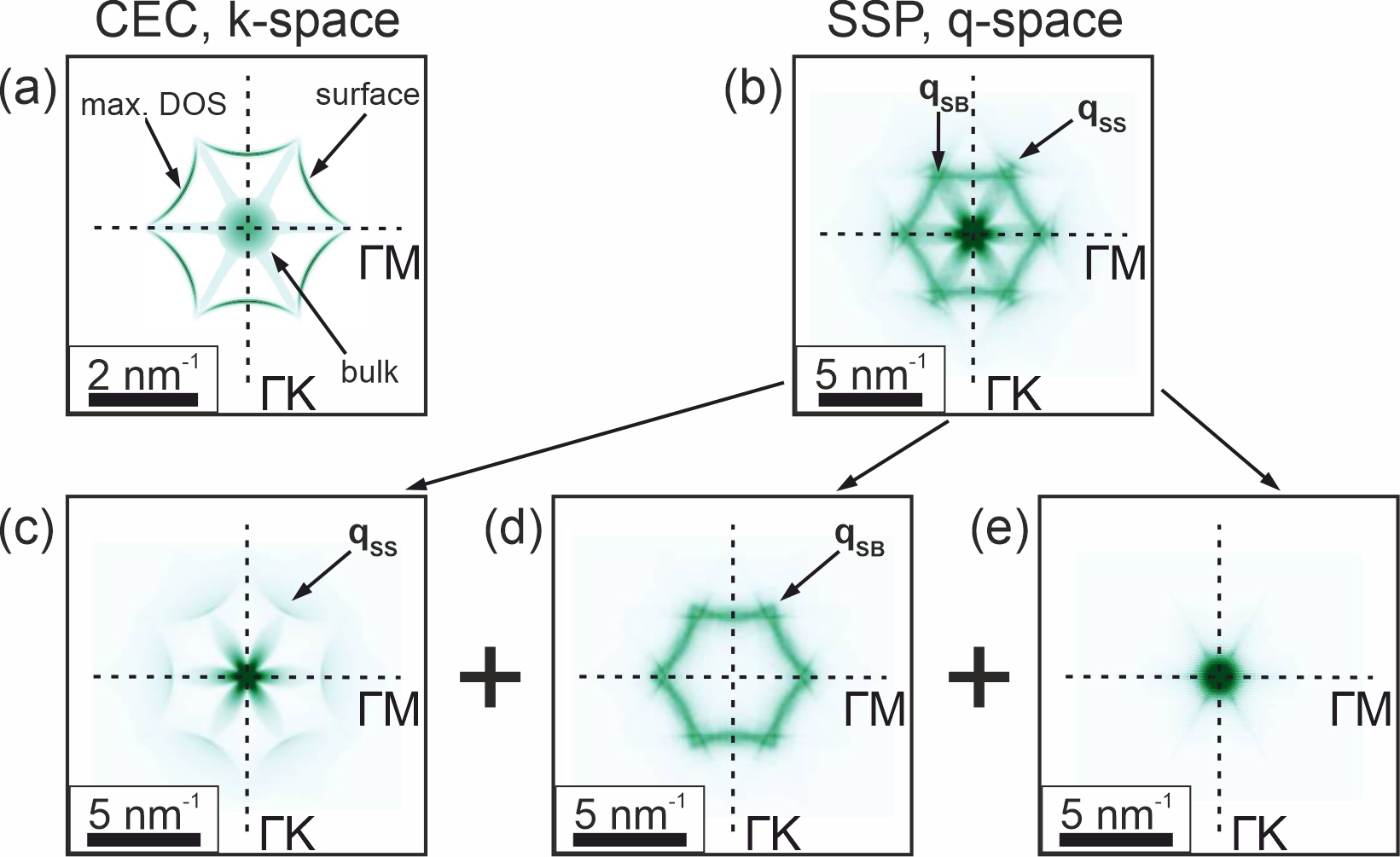}
		\caption{Different contributions to the total SSP pattern in (b) for the CEC plotted in (a) corresponding to the bulk conduction band. In particular, SSP calculations results for the surface-to-surface, bulk-to-surface and bulk-to-bulk scattering are presented in (c), (d) and (e), respectively. The color scale is the same as that in Fig.~\ref{fig:Fig.17}.}
		\label{fig:Fig.19}
	\end{figure}

	By analyzing the FT-QPI data measured in the energy range at which the bulk conduction band appears, it is important to distinguish contributions of the surface states and the bulk states into the SSP pattern (see the CEC in Fig.~\ref{fig:Fig.19}{\color{blue}{(a)}}). Fig.~\ref{fig:Fig.19}{\color{blue}{(b)}} represents a superposition of 3 SSP patterns arising from scattering only within the surface CEC, bulk-to-surface scattering and scattering only within the bulk conduction band, which are plotted separately in Figs.~\ref{fig:Fig.19}{\color{blue}{(c-e)}}, respectively.  
	
	\section*{APPENDIX C: MAGNETIC FIELD QPI DATA }
	\label{C}
	\renewcommand{\theequation}{C\arabic{equation}}	
	
\begin{figure}[H]
		\centering
		\includegraphics[scale=0.7]{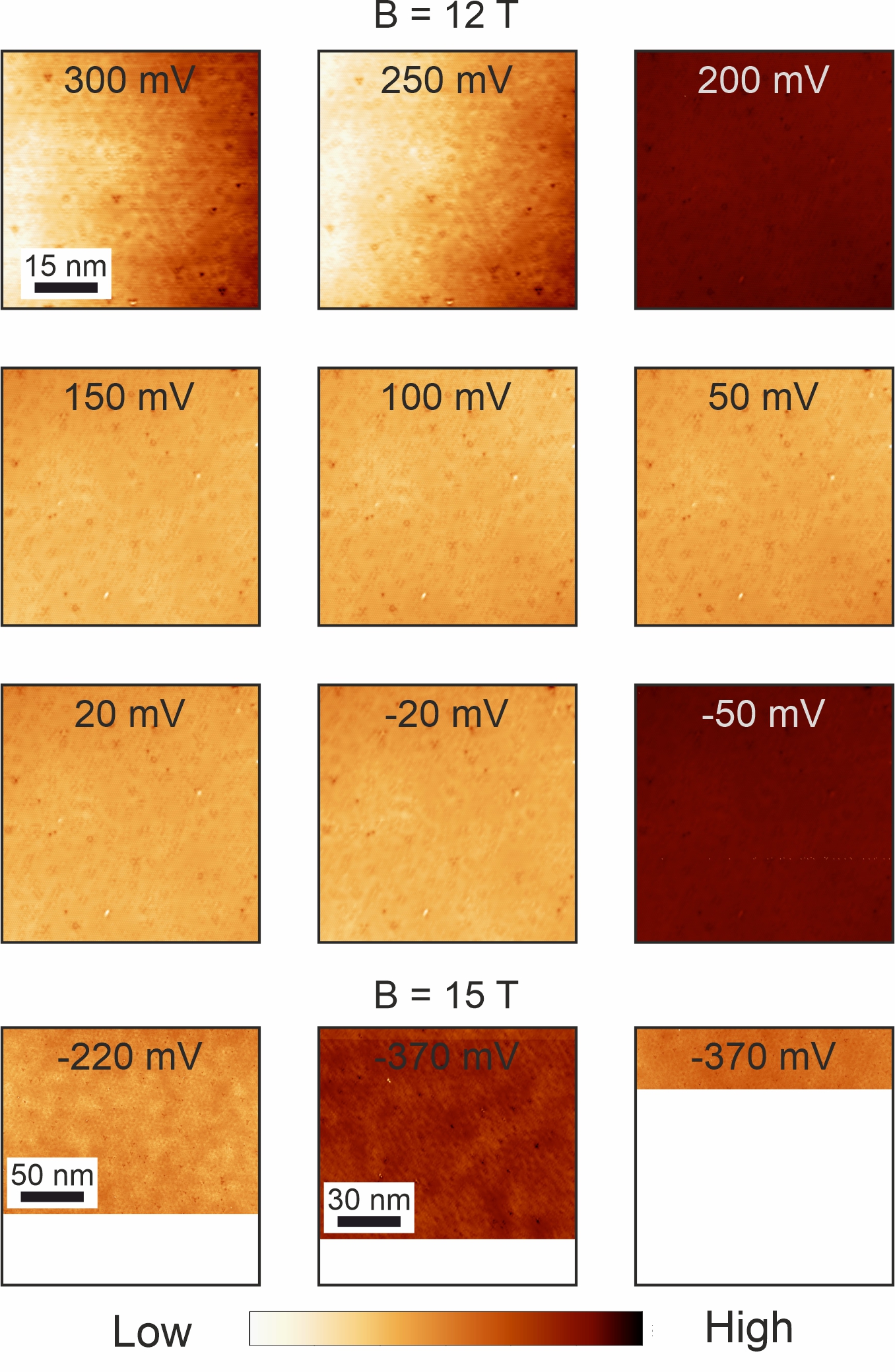}
		\caption{Series of the $dI/dU$ maps measured at $B = 12$ T and $B = 15$ T at the energies from 300 mV to $-50$ mV at $B = 12$ T and at $-220$ mV and $-370$ mV at $B = 15$ T. The scalebar is the same for each of the maps at $B = 12$ T. The scalebar for the map measured at $-370$ mV (shown in the right-bottom corner) is the same as that at $-220$ mV.}
		\label{fig:Fig.20}
	\end{figure}

\begin{figure}[]
		\centering
		\includegraphics[scale=0.46]{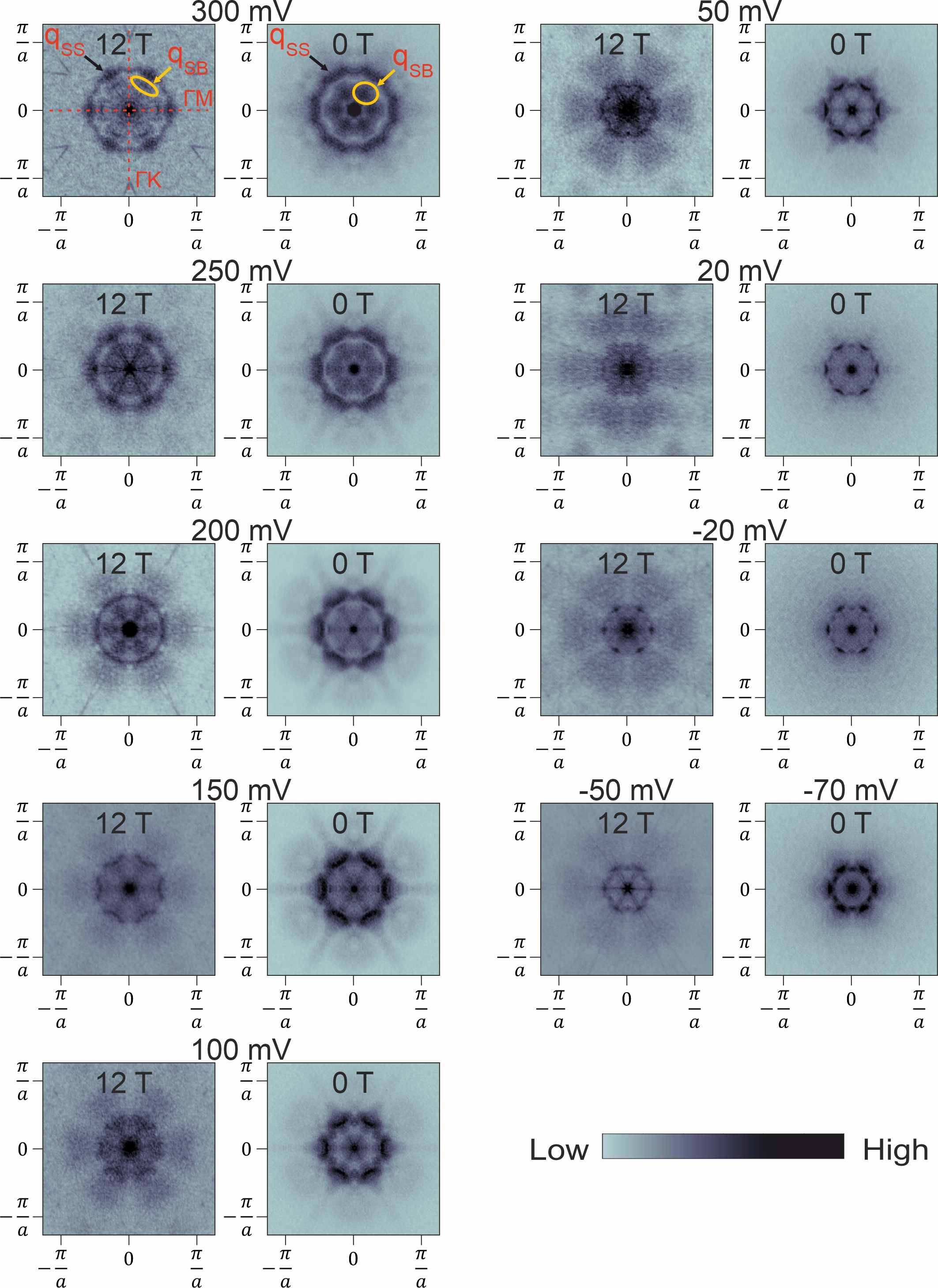}
		\caption{Series of the symmetrized FT-QPI patterns measured at the energies from 300 mV to $-50$ mV at $B = 12$ T and from 300 mV to $-70$ mV at $B = 0$ T. The outer intensity spots correspond to the vector \textbf{q}$_{SS}$ for each of the patterns. The inner intensity spot, which is related to the scattering vector \textbf{q}$_{SB}$, is visible only at 300 mV, 250 mV and 200 mV and is indicated by the yellow ellipse. High-symmetry directions and the colorbar are the same for each of the patterns.}  
		\label{fig:Fig.21}
	\end{figure}

	A dataset of the $dI/dU$ maps measured at different energies in magnetic fields of 12 T and 15 T is displayed in Fig.~\ref{fig:Fig.20}. The differential conduction maps were measured at $B = 15$ T on larger surface areas than the data at $B = 12$ T, which provides a higher resolution in the FT-QPI patterns at $B = 15$ T. The corresponding FT-QPI patterns for the $B = 12$ T data are plotted in Fig.~\ref{fig:Fig.21} along with the zero-field data at respective energies providing qualitatively similar results. Note that the presented data at $B = 12$ T were acquired on a different sample with a different tip than the data at $B = 0$ T, which can explain slightly different FT-QPI intensity distributions at a given energy.
	
\begin{figure}[H]
		\centering
		\includegraphics[scale=0.7]{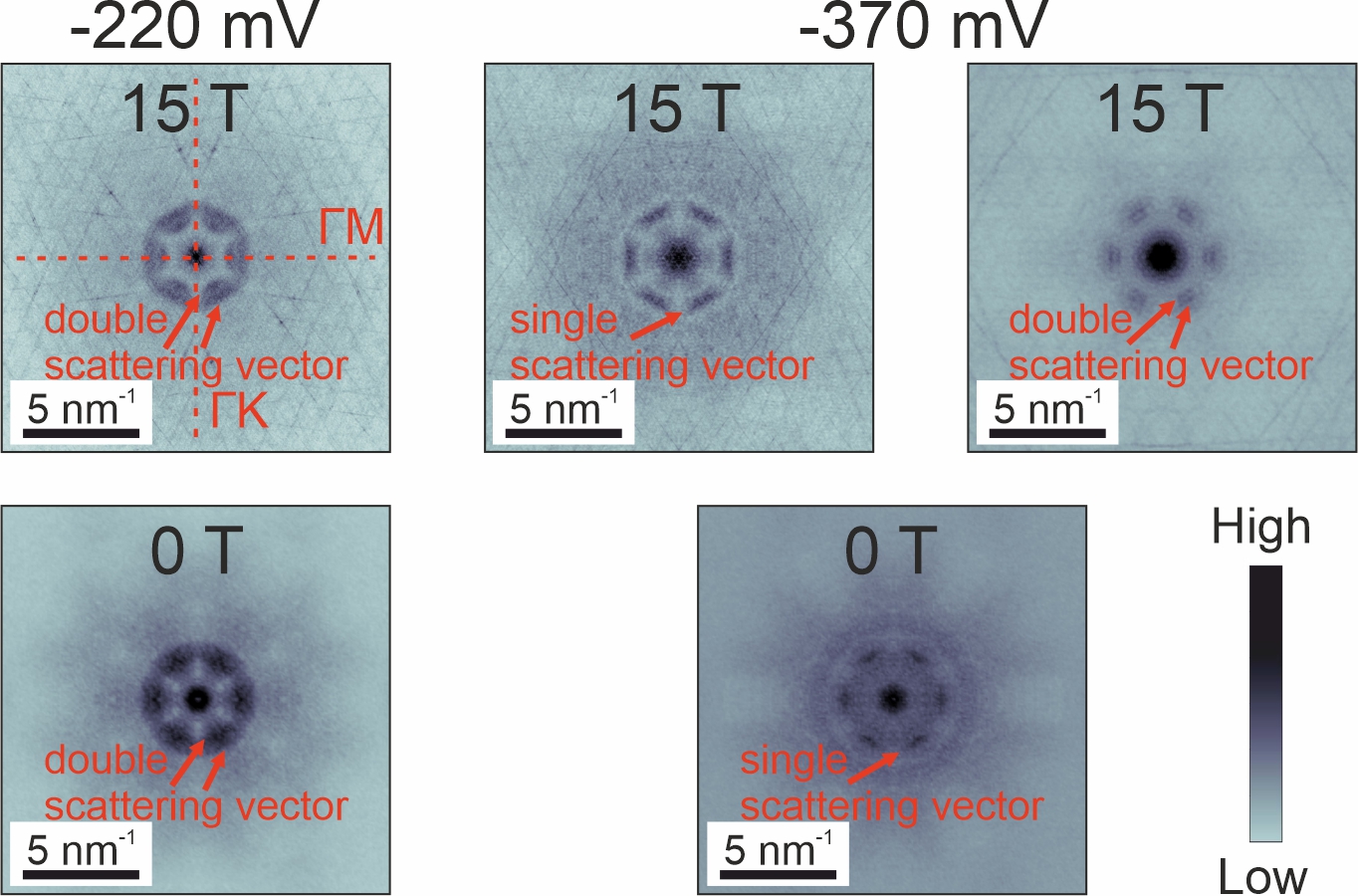}
		\caption{Symmetrized FT-QPI patterns measured at \\$U_{bias}$ = $-220$ mV and $U_{bias} = -370$ mV at $B = 15$ T in comparison with the data measured at $B = 0$ T at the same energies. The colorbar is the same for each of the patterns.}
		\label{fig:Fig.22}
	\end{figure}

	The FT-QPI patterns obtained at $B = 15$ T on the surface areas shown in Fig.~\ref{fig:Fig.20} are plotted in Fig.~\ref{fig:Fig.22} with the corresponding data measured at $B = 0$ T. The only strong difference is that there are two different FT-QPI patterns at $U_{bias} = -370$ mV at $B = 15$ T (exhibiting single and double dominant scattering vectors) which were measured on the same cleavage but at different surface areas indicating possible spatial variations of the Dirac point energy.     
	\newline


\begin{thebibliography}{40}%
\makeatletter
\providecommand \@ifxundefined [1]{%
 \@ifx{#1\undefined}
}%
\providecommand \@ifnum [1]{%
 \ifnum #1\expandafter \@firstoftwo
 \else \expandafter \@secondoftwo
 \fi
}%
\providecommand \@ifx [1]{%
 \ifx #1\expandafter \@firstoftwo
 \else \expandafter \@secondoftwo
 \fi
}%
\providecommand \natexlab [1]{#1}%
\providecommand \enquote  [1]{``#1''}%
\providecommand \bibnamefont  [1]{#1}%
\providecommand \bibfnamefont [1]{#1}%
\providecommand \citenamefont [1]{#1}%
\providecommand \href@noop [0]{\@secondoftwo}%
\providecommand \href [0]{\begingroup \@sanitize@url \@href}%
\providecommand \@href[1]{\@@startlink{#1}\@@href}%
\providecommand \@@href[1]{\endgroup#1\@@endlink}%
\providecommand \@sanitize@url [0]{\catcode `\\12\catcode `\$12\catcode
  `\&12\catcode `\#12\catcode `\^12\catcode `\_12\catcode `\%12\relax}%
\providecommand \@@startlink[1]{}%
\providecommand \@@endlink[0]{}%
\providecommand \url  [0]{\begingroup\@sanitize@url \@url }%
\providecommand \@url [1]{\endgroup\@href {#1}{\urlprefix }}%
\providecommand \urlprefix  [0]{URL }%
\providecommand \Eprint [0]{\href }%
\providecommand \doibase [0]{https://doi.org/}%
\providecommand \selectlanguage [0]{\@gobble}%
\providecommand \bibinfo  [0]{\@secondoftwo}%
\providecommand \bibfield  [0]{\@secondoftwo}%
\providecommand \translation [1]{[#1]}%
\providecommand \BibitemOpen [0]{}%
\providecommand \bibitemStop [0]{}%
\providecommand \bibitemNoStop [0]{.\EOS\space}%
\providecommand \EOS [0]{\spacefactor3000\relax}%
\providecommand \BibitemShut  [1]{\csname bibitem#1\endcsname}%
\let\auto@bib@innerbib\@empty
\bibitem [{\citenamefont {Fu}\ \emph {et~al.}(2007)\citenamefont {Fu},
  \citenamefont {Kane},\ and\ \citenamefont {Mele}}]{PhysRevLett.98.106803}%
  \BibitemOpen
  \bibfield  {author} {\bibinfo {author} {\bibfnamefont {L.}~\bibnamefont
  {Fu}}, \bibinfo {author} {\bibfnamefont {C.~L.}\ \bibnamefont {Kane}},\ and\
  \bibinfo {author} {\bibfnamefont {E.~J.}\ \bibnamefont {Mele}},\ }\bibfield
  {title} {\bibinfo {title} {Topological $\mathrm{I}$nsulators in
  $\mathrm{T}$hree $\mathrm{D}$imensions},\ }\href
  {https://doi.org/10.1103/PhysRevLett.98.106803} {\bibfield  {journal}
  {\bibinfo  {journal} {Phys. Rev. Lett.}\ }\textbf {\bibinfo {volume} {98}},\
  \bibinfo {pages} {106803} (\bibinfo {year} {2007})}\BibitemShut {NoStop}%
\bibitem [{\citenamefont {Moore}\ and\ \citenamefont
  {Balents}(2007)}]{PhysRevB.75.121306}%
  \BibitemOpen
  \bibfield  {author} {\bibinfo {author} {\bibfnamefont {J.~E.}\ \bibnamefont
  {Moore}}\ and\ \bibinfo {author} {\bibfnamefont {L.}~\bibnamefont
  {Balents}},\ }\bibfield  {title} {\bibinfo {title} {Topological invariants of
  time-reversal-invariant band structures},\ }\href
  {https://doi.org/10.1103/PhysRevB.75.121306} {\bibfield  {journal} {\bibinfo
  {journal} {Phys. Rev. B}\ }\textbf {\bibinfo {volume} {75}},\ \bibinfo
  {pages} {121306} (\bibinfo {year} {2007})}\BibitemShut {NoStop}%
\bibitem [{\citenamefont {Moore}(2009)}]{Moore2009}%
  \BibitemOpen
  \bibfield  {author} {\bibinfo {author} {\bibfnamefont {J.}~\bibnamefont
  {Moore}},\ }\bibfield  {title} {\bibinfo {title} {The next generation},\
  }\href {https://doi.org/10.1038/nphys1294} {\bibfield  {journal} {\bibinfo
  {journal} {Nature Physics}\ }\textbf {\bibinfo {volume} {5}},\ \bibinfo
  {pages} {378} (\bibinfo {year} {2009})}\BibitemShut {NoStop}%
\bibitem [{\citenamefont {Zhang}\ \emph
  {et~al.}(2009{\natexlab{a}})\citenamefont {Zhang}, \citenamefont {Liu},
  \citenamefont {Qi}, \citenamefont {Dai}, \citenamefont {Fang},\ and\
  \citenamefont {Zhang}}]{Zhang2009}%
  \BibitemOpen
  \bibfield  {author} {\bibinfo {author} {\bibfnamefont {H.}~\bibnamefont
  {Zhang}}, \bibinfo {author} {\bibfnamefont {C.-X.}\ \bibnamefont {Liu}},
  \bibinfo {author} {\bibfnamefont {X.-L.}\ \bibnamefont {Qi}}, \bibinfo
  {author} {\bibfnamefont {X.}~\bibnamefont {Dai}}, \bibinfo {author}
  {\bibfnamefont {Z.}~\bibnamefont {Fang}},\ and\ \bibinfo {author}
  {\bibfnamefont {S.-C.}\ \bibnamefont {Zhang}},\ }\bibfield  {title} {\bibinfo
  {title} {Topological insulators in
  $\mathrm{B}{\mathrm{i}}_{2}\mathrm{S}{\mathrm{e}}_{3}$,
  $\mathrm{B}{\mathrm{i}}_{2}\mathrm{T}{\mathrm{e}}_{3}$ and
  $\mathrm{S}{\mathrm{b}}_{2}\mathrm{T}{\mathrm{e}}_{3}$ with a single
  $\mathrm{D}$irac cone on the surface},\ }\href
  {https://doi.org/10.1038/nphys1270} {\bibfield  {journal} {\bibinfo
  {journal} {Nature Physics}\ }\textbf {\bibinfo {volume} {5}},\ \bibinfo
  {pages} {438} (\bibinfo {year} {2009}{\natexlab{a}})}\BibitemShut {NoStop}%
\bibitem [{\citenamefont {Xia}\ \emph {et~al.}(2009)\citenamefont {Xia},
  \citenamefont {Qian}, \citenamefont {Hsieh}, \citenamefont {Wray},
  \citenamefont {Pal}, \citenamefont {Lin}, \citenamefont {Bansil},
  \citenamefont {Grauer}, \citenamefont {Hor}, \citenamefont {Cava},\ and\
  \citenamefont {Hasan}}]{Xia2009}%
  \BibitemOpen
  \bibfield  {author} {\bibinfo {author} {\bibfnamefont {Y.}~\bibnamefont
  {Xia}}, \bibinfo {author} {\bibfnamefont {D.}~\bibnamefont {Qian}}, \bibinfo
  {author} {\bibfnamefont {D.}~\bibnamefont {Hsieh}}, \bibinfo {author}
  {\bibfnamefont {L.}~\bibnamefont {Wray}}, \bibinfo {author} {\bibfnamefont
  {A.}~\bibnamefont {Pal}}, \bibinfo {author} {\bibfnamefont {H.}~\bibnamefont
  {Lin}}, \bibinfo {author} {\bibfnamefont {A.}~\bibnamefont {Bansil}},
  \bibinfo {author} {\bibfnamefont {D.}~\bibnamefont {Grauer}}, \bibinfo
  {author} {\bibfnamefont {Y.~S.}\ \bibnamefont {Hor}}, \bibinfo {author}
  {\bibfnamefont {R.~J.}\ \bibnamefont {Cava}},\ and\ \bibinfo {author}
  {\bibfnamefont {M.~Z.}\ \bibnamefont {Hasan}},\ }\bibfield  {title} {\bibinfo
  {title} {Observation of a large-gap topological-insulator class with a single
  $\mathrm{D}$irac cone on the surface},\ }\href
  {https://doi.org/10.1038/nphys1274} {\bibfield  {journal} {\bibinfo
  {journal} {Nature Physics}\ }\textbf {\bibinfo {volume} {5}},\ \bibinfo
  {pages} {398} (\bibinfo {year} {2009})}\BibitemShut {NoStop}%
\bibitem [{\citenamefont {Chen}\ \emph {et~al.}(2009)\citenamefont {Chen},
  \citenamefont {Analytis}, \citenamefont {Chu}, \citenamefont {Liu},
  \citenamefont {Mo}, \citenamefont {Qi}, \citenamefont {Zhang}, \citenamefont
  {Lu}, \citenamefont {Dai}, \citenamefont {Fang}, \citenamefont {Zhang},
  \citenamefont {Fisher}, \citenamefont {Hussain},\ and\ \citenamefont
  {Shen}}]{doi:10.1126/science.1173034}%
  \BibitemOpen
  \bibfield  {author} {\bibinfo {author} {\bibfnamefont {Y.~L.}\ \bibnamefont
  {Chen}}, \bibinfo {author} {\bibfnamefont {J.~G.}\ \bibnamefont {Analytis}},
  \bibinfo {author} {\bibfnamefont {J.-H.}\ \bibnamefont {Chu}}, \bibinfo
  {author} {\bibfnamefont {Z.~K.}\ \bibnamefont {Liu}}, \bibinfo {author}
  {\bibfnamefont {S.-K.}\ \bibnamefont {Mo}}, \bibinfo {author} {\bibfnamefont
  {X.~L.}\ \bibnamefont {Qi}}, \bibinfo {author} {\bibfnamefont {H.~J.}\
  \bibnamefont {Zhang}}, \bibinfo {author} {\bibfnamefont {D.~H.}\ \bibnamefont
  {Lu}}, \bibinfo {author} {\bibfnamefont {X.}~\bibnamefont {Dai}}, \bibinfo
  {author} {\bibfnamefont {Z.}~\bibnamefont {Fang}}, \bibinfo {author}
  {\bibfnamefont {S.~C.}\ \bibnamefont {Zhang}}, \bibinfo {author}
  {\bibfnamefont {I.~R.}\ \bibnamefont {Fisher}}, \bibinfo {author}
  {\bibfnamefont {Z.}~\bibnamefont {Hussain}},\ and\ \bibinfo {author}
  {\bibfnamefont {Z.-X.}\ \bibnamefont {Shen}},\ }\bibfield  {title} {\bibinfo
  {title} {Experimental $\mathrm{R}$ealization of a
  $\mathrm{T}$hree-$\mathrm{D}$imensional $\mathrm{T}$opological
  $\mathrm{I}$nsulator,
  $\mathrm{B}{\mathrm{i}}_{2}\mathrm{T}{\mathrm{e}}_{3}$},\ }\href
  {https://doi.org/10.1126/science.1173034} {\bibfield  {journal} {\bibinfo
  {journal} {Science}\ }\textbf {\bibinfo {volume} {325}},\ \bibinfo {pages}
  {178} (\bibinfo {year} {2009})}\BibitemShut {NoStop}%
\bibitem [{\citenamefont {Hasan}\ and\ \citenamefont
  {Kane}(2010)}]{RevModPhys.82.3045}%
  \BibitemOpen
  \bibfield  {author} {\bibinfo {author} {\bibfnamefont {M.~Z.}\ \bibnamefont
  {Hasan}}\ and\ \bibinfo {author} {\bibfnamefont {C.~L.}\ \bibnamefont
  {Kane}},\ }\bibfield  {title} {\bibinfo {title} {Colloquium:
  $\mathrm{T}$opological insulators},\ }\href
  {https://doi.org/10.1103/RevModPhys.82.3045} {\bibfield  {journal} {\bibinfo
  {journal} {Rev. Mod. Phys.}\ }\textbf {\bibinfo {volume} {82}},\ \bibinfo
  {pages} {3045} (\bibinfo {year} {2010})}\BibitemShut {NoStop}%
\bibitem [{\citenamefont {Fu}(2009)}]{PhysRevLett.103.266801}%
  \BibitemOpen
  \bibfield  {author} {\bibinfo {author} {\bibfnamefont {L.}~\bibnamefont
  {Fu}},\ }\bibfield  {title} {\bibinfo {title} {Hexagonal $\mathrm{W}$arping
  $\mathrm{E}$ffects in the $\mathrm{S}$urface $\mathrm{S}$tates of the
  $\mathrm{T}$opological $\mathrm{I}$nsulator
  $\mathrm{B}{\mathrm{i}}_{2}\mathrm{T}{\mathrm{e}}_{3}$},\ }\href
  {https://doi.org/10.1103/PhysRevLett.103.266801} {\bibfield  {journal}
  {\bibinfo  {journal} {Phys. Rev. Lett.}\ }\textbf {\bibinfo {volume} {103}},\
  \bibinfo {pages} {266801} (\bibinfo {year} {2009})}\BibitemShut {NoStop}%
\bibitem [{\citenamefont {Hasan}\ \emph {et~al.}(2009)\citenamefont {Hasan},
  \citenamefont {Lin},\ and\ \citenamefont {Bansil}}]{2009}%
  \BibitemOpen
  \bibfield  {author} {\bibinfo {author} {\bibfnamefont {M.~Z.}\ \bibnamefont
  {Hasan}}, \bibinfo {author} {\bibfnamefont {H.}~\bibnamefont {Lin}},\ and\
  \bibinfo {author} {\bibfnamefont {A.}~\bibnamefont {Bansil}},\ }\bibfield
  {title} {\bibinfo {title} {Warping the cone on a $\mathrm{T}$opological
  $\mathrm{I}$nsulator},\ }\href
  {https://physics.aps.org/articles/pdf/10.1103/Physics.2.108} {\bibfield
  {journal} {\bibinfo  {journal} {Physics}\ }\textbf {\bibinfo {volume} {108}}
  (\bibinfo {year} {2009})}\BibitemShut {NoStop}%
\bibitem [{\citenamefont {Sprunger}\ \emph {et~al.}(1997)\citenamefont
  {Sprunger}, \citenamefont {Petersen}, \citenamefont {Plummer}, \citenamefont
  {Lægsgaard},\ and\ \citenamefont {Besenbacher}}]{Sprunger1997}%
  \BibitemOpen
  \bibfield  {author} {\bibinfo {author} {\bibfnamefont {P.~T.}\ \bibnamefont
  {Sprunger}}, \bibinfo {author} {\bibfnamefont {L.}~\bibnamefont {Petersen}},
  \bibinfo {author} {\bibfnamefont {E.~W.}\ \bibnamefont {Plummer}}, \bibinfo
  {author} {\bibfnamefont {E.}~\bibnamefont {Lægsgaard}},\ and\ \bibinfo
  {author} {\bibfnamefont {F.}~\bibnamefont {Besenbacher}},\ }\bibfield
  {title} {\bibinfo {title} {Giant $\mathrm{F}$riedel $\mathrm{O}$scillations
  on the $\mathrm{B}$eryllium(0001) $\mathrm{S}$urface},\ }\href
  {https://doi.org/10.1126/science.275.5307.1764} {\bibfield  {journal}
  {\bibinfo  {journal} {Science}\ }\textbf {\bibinfo {volume} {275}},\ \bibinfo
  {pages} {1764} (\bibinfo {year} {1997})}\BibitemShut {NoStop}%
\bibitem [{\citenamefont {Petersen}\ \emph {et~al.}(1998)\citenamefont
  {Petersen}, \citenamefont {Sprunger}, \citenamefont {Hofmann}, \citenamefont
  {L\ae{}gsgaard}, \citenamefont {Briner}, \citenamefont {Doering},
  \citenamefont {Rust}, \citenamefont {Bradshaw}, \citenamefont {Besenbacher},\
  and\ \citenamefont {Plummer}}]{PhysRevB.57.R6858}%
  \BibitemOpen
  \bibfield  {author} {\bibinfo {author} {\bibfnamefont {L.}~\bibnamefont
  {Petersen}}, \bibinfo {author} {\bibfnamefont {P.~T.}\ \bibnamefont
  {Sprunger}}, \bibinfo {author} {\bibfnamefont {P.}~\bibnamefont {Hofmann}},
  \bibinfo {author} {\bibfnamefont {E.}~\bibnamefont {L\ae{}gsgaard}}, \bibinfo
  {author} {\bibfnamefont {B.~G.}\ \bibnamefont {Briner}}, \bibinfo {author}
  {\bibfnamefont {M.}~\bibnamefont {Doering}}, \bibinfo {author} {\bibfnamefont
  {H.-P.}\ \bibnamefont {Rust}}, \bibinfo {author} {\bibfnamefont {A.~M.}\
  \bibnamefont {Bradshaw}}, \bibinfo {author} {\bibfnamefont {F.}~\bibnamefont
  {Besenbacher}},\ and\ \bibinfo {author} {\bibfnamefont {E.~W.}\ \bibnamefont
  {Plummer}},\ }\bibfield  {title} {\bibinfo {title} {Direct imaging of the
  two-dimensional $\mathrm{F}$ermi contour: $\mathrm{F}$ourier-transform
  $\mathrm{STM}$},\ }\href {https://doi.org/10.1103/PhysRevB.57.R6858}
  {\bibfield  {journal} {\bibinfo  {journal} {Phys. Rev. B}\ }\textbf {\bibinfo
  {volume} {57}},\ \bibinfo {pages} {R6858} (\bibinfo {year}
  {1998})}\BibitemShut {NoStop}%
\bibitem [{\citenamefont {Hoffman}\ \emph {et~al.}(2002)\citenamefont
  {Hoffman}, \citenamefont {McElroy}, \citenamefont {Lee}, \citenamefont
  {Lang}, \citenamefont {Eisaki}, \citenamefont {Uchida},\ and\ \citenamefont
  {Davis}}]{Hoffman2002}%
  \BibitemOpen
  \bibfield  {author} {\bibinfo {author} {\bibfnamefont {J.~E.}\ \bibnamefont
  {Hoffman}}, \bibinfo {author} {\bibfnamefont {K.}~\bibnamefont {McElroy}},
  \bibinfo {author} {\bibfnamefont {D.-H.}\ \bibnamefont {Lee}}, \bibinfo
  {author} {\bibfnamefont {K.~M.}\ \bibnamefont {Lang}}, \bibinfo {author}
  {\bibfnamefont {H.}~\bibnamefont {Eisaki}}, \bibinfo {author} {\bibfnamefont
  {S.}~\bibnamefont {Uchida}},\ and\ \bibinfo {author} {\bibfnamefont {J.~C.}\
  \bibnamefont {Davis}},\ }\bibfield  {title} {\bibinfo {title} {Imaging
  quasiparticle interference in
  $\mathrm{B}{\mathrm{i}}_{2}\mathrm{S}{\mathrm{r}}_{2}\mathrm{C}{\mathrm{a}}\mathrm{C}{\mathrm{u}}_{2}\mathrm{O}_{8+\delta}$},\
  }\href {https://doi.org/10.1126/science.1072640} {\bibfield  {journal}
  {\bibinfo  {journal} {Science}\ }\textbf {\bibinfo {volume} {297}},\ \bibinfo
  {pages} {1148} (\bibinfo {year} {2002})}\BibitemShut {NoStop}%
\bibitem [{\citenamefont {Schmidt}\ \emph {et~al.}(2010)\citenamefont
  {Schmidt}, \citenamefont {Hamidian}, \citenamefont {Wahl}, \citenamefont
  {Meier}, \citenamefont {Balatsky}, \citenamefont {Garrett}, \citenamefont
  {Williams}, \citenamefont {Luke},\ and\ \citenamefont {Davis}}]{Schmidt2010}%
  \BibitemOpen
  \bibfield  {author} {\bibinfo {author} {\bibfnamefont {A.~R.}\ \bibnamefont
  {Schmidt}}, \bibinfo {author} {\bibfnamefont {M.~H.}\ \bibnamefont
  {Hamidian}}, \bibinfo {author} {\bibfnamefont {P.}~\bibnamefont {Wahl}},
  \bibinfo {author} {\bibfnamefont {F.}~\bibnamefont {Meier}}, \bibinfo
  {author} {\bibfnamefont {A.~V.}\ \bibnamefont {Balatsky}}, \bibinfo {author}
  {\bibfnamefont {J.~D.}\ \bibnamefont {Garrett}}, \bibinfo {author}
  {\bibfnamefont {T.~J.}\ \bibnamefont {Williams}}, \bibinfo {author}
  {\bibfnamefont {G.~M.}\ \bibnamefont {Luke}},\ and\ \bibinfo {author}
  {\bibfnamefont {J.~C.}\ \bibnamefont {Davis}},\ }\bibfield  {title} {\bibinfo
  {title} {Imaging the $\mathrm{F}$ano lattice to ‘hidden order’ transition
  in $\mathrm{U}$$\mathrm{R}{\mathrm{u}}_{2}\mathrm{S}{\mathrm{i}}_{2}$},\
  }\href {https://doi.org/10.1038/nature09073} {\bibfield  {journal} {\bibinfo
  {journal} {Nature}\ }\textbf {\bibinfo {volume} {465}},\ \bibinfo {pages}
  {570} (\bibinfo {year} {2010})}\BibitemShut {NoStop}%
\bibitem [{\citenamefont {Aynajian}\ \emph {et~al.}(2012)\citenamefont
  {Aynajian}, \citenamefont {da~Silva~Neto}, \citenamefont {Gyenis},
  \citenamefont {Baumbach}, \citenamefont {Thompson}, \citenamefont {Fisk},
  \citenamefont {Bauer},\ and\ \citenamefont {Yazdani}}]{Aynajian2012}%
  \BibitemOpen
  \bibfield  {author} {\bibinfo {author} {\bibfnamefont {P.}~\bibnamefont
  {Aynajian}}, \bibinfo {author} {\bibfnamefont {E.~H.}\ \bibnamefont
  {da~Silva~Neto}}, \bibinfo {author} {\bibfnamefont {A.}~\bibnamefont
  {Gyenis}}, \bibinfo {author} {\bibfnamefont {R.~E.}\ \bibnamefont
  {Baumbach}}, \bibinfo {author} {\bibfnamefont {J.~D.}\ \bibnamefont
  {Thompson}}, \bibinfo {author} {\bibfnamefont {Z.}~\bibnamefont {Fisk}},
  \bibinfo {author} {\bibfnamefont {E.~D.}\ \bibnamefont {Bauer}},\ and\
  \bibinfo {author} {\bibfnamefont {A.}~\bibnamefont {Yazdani}},\ }\bibfield
  {title} {\bibinfo {title} {Visualizing heavy fermions emerging in a quantum
  critical $\mathrm{K}$ondo lattice},\ }\href
  {https://doi.org/10.1038/nature11204} {\bibfield  {journal} {\bibinfo
  {journal} {Nature}\ }\textbf {\bibinfo {volume} {486}},\ \bibinfo {pages}
  {201} (\bibinfo {year} {2012})}\BibitemShut {NoStop}%
\bibitem [{\citenamefont {Zheng}\ \emph {et~al.}(2016)\citenamefont {Zheng},
  \citenamefont {Xu}, \citenamefont {Bian}, \citenamefont {Guo}, \citenamefont
  {Chang}, \citenamefont {Sanchez}, \citenamefont {Belopolski}, \citenamefont
  {Lee}, \citenamefont {Huang}, \citenamefont {Zhang}, \citenamefont {Sankar},
  \citenamefont {Alidoust}, \citenamefont {Chang}, \citenamefont {Wu},
  \citenamefont {Neupert}, \citenamefont {Chou}, \citenamefont {Jeng},
  \citenamefont {Yao}, \citenamefont {Bansil}, \citenamefont {Jia},
  \citenamefont {Lin},\ and\ \citenamefont {Hasan}}]{Zheng2016}%
  \BibitemOpen
  \bibfield  {author} {\bibinfo {author} {\bibfnamefont {H.}~\bibnamefont
  {Zheng}}, \bibinfo {author} {\bibfnamefont {S.-Y.}\ \bibnamefont {Xu}},
  \bibinfo {author} {\bibfnamefont {G.}~\bibnamefont {Bian}}, \bibinfo {author}
  {\bibfnamefont {C.}~\bibnamefont {Guo}}, \bibinfo {author} {\bibfnamefont
  {G.}~\bibnamefont {Chang}}, \bibinfo {author} {\bibfnamefont {D.~S.}\
  \bibnamefont {Sanchez}}, \bibinfo {author} {\bibfnamefont {I.}~\bibnamefont
  {Belopolski}}, \bibinfo {author} {\bibfnamefont {C.-C.}\ \bibnamefont {Lee}},
  \bibinfo {author} {\bibfnamefont {S.-M.}\ \bibnamefont {Huang}}, \bibinfo
  {author} {\bibfnamefont {X.}~\bibnamefont {Zhang}}, \bibinfo {author}
  {\bibfnamefont {R.}~\bibnamefont {Sankar}}, \bibinfo {author} {\bibfnamefont
  {N.}~\bibnamefont {Alidoust}}, \bibinfo {author} {\bibfnamefont {T.-R.}\
  \bibnamefont {Chang}}, \bibinfo {author} {\bibfnamefont {F.}~\bibnamefont
  {Wu}}, \bibinfo {author} {\bibfnamefont {T.}~\bibnamefont {Neupert}},
  \bibinfo {author} {\bibfnamefont {F.}~\bibnamefont {Chou}}, \bibinfo {author}
  {\bibfnamefont {H.-T.}\ \bibnamefont {Jeng}}, \bibinfo {author}
  {\bibfnamefont {N.}~\bibnamefont {Yao}}, \bibinfo {author} {\bibfnamefont
  {A.}~\bibnamefont {Bansil}}, \bibinfo {author} {\bibfnamefont
  {S.}~\bibnamefont {Jia}}, \bibinfo {author} {\bibfnamefont {H.}~\bibnamefont
  {Lin}},\ and\ \bibinfo {author} {\bibfnamefont {M.~Z.}\ \bibnamefont
  {Hasan}},\ }\bibfield  {title} {\bibinfo {title} {Atomic-$\mathrm{S}$cale
  $\mathrm{V}$isualization of $\mathrm{Q}$uantum $\mathrm{I}$nterference on a
  $\mathrm{W}$eyl $\mathrm{S}$emimetal $\mathrm{S}$urface by
  $\mathrm{S}$canning $\mathrm{T}$unneling $\mathrm{M}$icroscopy},\ }\href
  {https://doi.org/10.1021/acsnano.5b06807} {\bibfield  {journal} {\bibinfo
  {journal} {ACS Nano}\ }\textbf {\bibinfo {volume} {10}},\ \bibinfo {pages}
  {1378} (\bibinfo {year} {2016})}\BibitemShut {NoStop}%
\bibitem [{\citenamefont {Inoue}\ \emph {et~al.}(2016)\citenamefont {Inoue},
  \citenamefont {Gyenis}, \citenamefont {Wang}, \citenamefont {Li},
  \citenamefont {Oh}, \citenamefont {Jiang}, \citenamefont {Ni}, \citenamefont
  {Bernevig},\ and\ \citenamefont {Yazdani}}]{Inoue2016}%
  \BibitemOpen
  \bibfield  {author} {\bibinfo {author} {\bibfnamefont {H.}~\bibnamefont
  {Inoue}}, \bibinfo {author} {\bibfnamefont {A.}~\bibnamefont {Gyenis}},
  \bibinfo {author} {\bibfnamefont {Z.}~\bibnamefont {Wang}}, \bibinfo {author}
  {\bibfnamefont {J.}~\bibnamefont {Li}}, \bibinfo {author} {\bibfnamefont
  {S.~W.}\ \bibnamefont {Oh}}, \bibinfo {author} {\bibfnamefont
  {S.}~\bibnamefont {Jiang}}, \bibinfo {author} {\bibfnamefont
  {N.}~\bibnamefont {Ni}}, \bibinfo {author} {\bibfnamefont {B.~A.}\
  \bibnamefont {Bernevig}},\ and\ \bibinfo {author} {\bibfnamefont
  {A.}~\bibnamefont {Yazdani}},\ }\bibfield  {title} {\bibinfo {title}
  {Quasiparticle interference of the $\mathrm{F}$ermi arcs and surface-bulk
  connectivity of a $\mathrm{W}$eyl semimetal},\ }\href
  {https://doi.org/10.1126/science.aad8766} {\bibfield  {journal} {\bibinfo
  {journal} {Science}\ }\textbf {\bibinfo {volume} {351}},\ \bibinfo {pages}
  {1184} (\bibinfo {year} {2016})}\BibitemShut {NoStop}%
\bibitem [{\citenamefont {Roushan}\ \emph {et~al.}(2009)\citenamefont
  {Roushan}, \citenamefont {Seo}, \citenamefont {Parker}, \citenamefont {Hor},
  \citenamefont {Hsieh}, \citenamefont {Qian}, \citenamefont {Richardella},
  \citenamefont {Hasan}, \citenamefont {Cava},\ and\ \citenamefont
  {Yazdani}}]{Roushan2009}%
  \BibitemOpen
  \bibfield  {author} {\bibinfo {author} {\bibfnamefont {P.}~\bibnamefont
  {Roushan}}, \bibinfo {author} {\bibfnamefont {J.}~\bibnamefont {Seo}},
  \bibinfo {author} {\bibfnamefont {C.~V.}\ \bibnamefont {Parker}}, \bibinfo
  {author} {\bibfnamefont {Y.~S.}\ \bibnamefont {Hor}}, \bibinfo {author}
  {\bibfnamefont {D.}~\bibnamefont {Hsieh}}, \bibinfo {author} {\bibfnamefont
  {D.}~\bibnamefont {Qian}}, \bibinfo {author} {\bibfnamefont {A.}~\bibnamefont
  {Richardella}}, \bibinfo {author} {\bibfnamefont {M.~Z.}\ \bibnamefont
  {Hasan}}, \bibinfo {author} {\bibfnamefont {R.~J.}\ \bibnamefont {Cava}},\
  and\ \bibinfo {author} {\bibfnamefont {A.}~\bibnamefont {Yazdani}},\
  }\bibfield  {title} {\bibinfo {title} {Topological surface states protected
  from backscattering by chiral spin texture},\ }\href
  {https://doi.org/10.1038/nature08308} {\bibfield  {journal} {\bibinfo
  {journal} {Nature}\ }\textbf {\bibinfo {volume} {460}},\ \bibinfo {pages}
  {1106} (\bibinfo {year} {2009})}\BibitemShut {NoStop}%
\bibitem [{\citenamefont {Zhang}\ \emph
  {et~al.}(2009{\natexlab{b}})\citenamefont {Zhang}, \citenamefont {Cheng},
  \citenamefont {Chen}, \citenamefont {Jia}, \citenamefont {Ma}, \citenamefont
  {He}, \citenamefont {Wang}, \citenamefont {Zhang}, \citenamefont {Dai},
  \citenamefont {Fang}, \citenamefont {Xie},\ and\ \citenamefont
  {Xue}}]{PhysRevLett.103.266803}%
  \BibitemOpen
  \bibfield  {author} {\bibinfo {author} {\bibfnamefont {T.}~\bibnamefont
  {Zhang}}, \bibinfo {author} {\bibfnamefont {P.}~\bibnamefont {Cheng}},
  \bibinfo {author} {\bibfnamefont {X.}~\bibnamefont {Chen}}, \bibinfo {author}
  {\bibfnamefont {J.-F.}\ \bibnamefont {Jia}}, \bibinfo {author} {\bibfnamefont
  {X.}~\bibnamefont {Ma}}, \bibinfo {author} {\bibfnamefont {K.}~\bibnamefont
  {He}}, \bibinfo {author} {\bibfnamefont {L.}~\bibnamefont {Wang}}, \bibinfo
  {author} {\bibfnamefont {H.}~\bibnamefont {Zhang}}, \bibinfo {author}
  {\bibfnamefont {X.}~\bibnamefont {Dai}}, \bibinfo {author} {\bibfnamefont
  {Z.}~\bibnamefont {Fang}}, \bibinfo {author} {\bibfnamefont {X.}~\bibnamefont
  {Xie}},\ and\ \bibinfo {author} {\bibfnamefont {Q.-K.}\ \bibnamefont {Xue}},\
  }\bibfield  {title} {\bibinfo {title} {Experimental $\mathrm{D}$emonstration
  of $\mathrm{T}$opological $\mathrm{S}$urface $\mathrm{S}$tates
  $\mathrm{P}$rotected by $\mathrm{T}$ime-$\mathrm{R}$eversal
  $\mathrm{S}$ymmetry},\ }\href
  {https://doi.org/10.1103/PhysRevLett.103.266803} {\bibfield  {journal}
  {\bibinfo  {journal} {Phys. Rev. Lett.}\ }\textbf {\bibinfo {volume} {103}},\
  \bibinfo {pages} {266803} (\bibinfo {year} {2009}{\natexlab{b}})}\BibitemShut
  {NoStop}%
\bibitem [{\citenamefont {Alpichshev}\ \emph {et~al.}(2010)\citenamefont
  {Alpichshev}, \citenamefont {Analytis}, \citenamefont {Chu}, \citenamefont
  {Fisher}, \citenamefont {Chen}, \citenamefont {Shen}, \citenamefont {Fang},\
  and\ \citenamefont {Kapitulnik}}]{PhysRevLett.104.016401}%
  \BibitemOpen
  \bibfield  {author} {\bibinfo {author} {\bibfnamefont {Z.}~\bibnamefont
  {Alpichshev}}, \bibinfo {author} {\bibfnamefont {J.~G.}\ \bibnamefont
  {Analytis}}, \bibinfo {author} {\bibfnamefont {J.-H.}\ \bibnamefont {Chu}},
  \bibinfo {author} {\bibfnamefont {I.~R.}\ \bibnamefont {Fisher}}, \bibinfo
  {author} {\bibfnamefont {Y.~L.}\ \bibnamefont {Chen}}, \bibinfo {author}
  {\bibfnamefont {Z.~X.}\ \bibnamefont {Shen}}, \bibinfo {author}
  {\bibfnamefont {A.}~\bibnamefont {Fang}},\ and\ \bibinfo {author}
  {\bibfnamefont {A.}~\bibnamefont {Kapitulnik}},\ }\bibfield  {title}
  {\bibinfo {title} {S$\mathrm{TM}$ $\mathrm{I}$maging of $\mathrm{E}$lectronic
  $\mathrm{W}$aves on the $\mathrm{S}$urface of
  $\mathrm{B}{\mathrm{i}}_{2}\mathrm{T}{\mathrm{e}}_{3}$:
  $\mathrm{T}$opologically $\mathrm{P}$rotected $\mathrm{S}$urface
  $\mathrm{S}$tates and $\mathrm{H}$exagonal $\mathrm{W}$arping
  $\mathrm{E}$ffects},\ }\href {https://doi.org/10.1103/PhysRevLett.104.016401}
  {\bibfield  {journal} {\bibinfo  {journal} {Phys. Rev. Lett.}\ }\textbf
  {\bibinfo {volume} {104}},\ \bibinfo {pages} {016401} (\bibinfo {year}
  {2010})}\BibitemShut {NoStop}%
\bibitem [{\citenamefont {Urazhdin}\ \emph {et~al.}(2004)\citenamefont
  {Urazhdin}, \citenamefont {Bilc}, \citenamefont {Mahanti}, \citenamefont
  {Tessmer}, \citenamefont {Kyratsi},\ and\ \citenamefont
  {Kanatzidis}}]{PhysRevB.69.085313}%
  \BibitemOpen
  \bibfield  {author} {\bibinfo {author} {\bibfnamefont {S.}~\bibnamefont
  {Urazhdin}}, \bibinfo {author} {\bibfnamefont {D.}~\bibnamefont {Bilc}},
  \bibinfo {author} {\bibfnamefont {S.~D.}\ \bibnamefont {Mahanti}}, \bibinfo
  {author} {\bibfnamefont {S.~H.}\ \bibnamefont {Tessmer}}, \bibinfo {author}
  {\bibfnamefont {T.}~\bibnamefont {Kyratsi}},\ and\ \bibinfo {author}
  {\bibfnamefont {M.~G.}\ \bibnamefont {Kanatzidis}},\ }\bibfield  {title}
  {\bibinfo {title} {Surface effects in layered semiconductors
  $\mathrm{B}{\mathrm{i}}_{2}\mathrm{S}{\mathrm{e}}_{3}$ and
  $\mathrm{B}{\mathrm{i}}_{2}\mathrm{T}{\mathrm{e}}_{3}$},\ }\href
  {https://doi.org/10.1103/PhysRevB.69.085313} {\bibfield  {journal} {\bibinfo
  {journal} {Phys. Rev. B}\ }\textbf {\bibinfo {volume} {69}},\ \bibinfo
  {pages} {085313} (\bibinfo {year} {2004})}\BibitemShut {NoStop}%
\bibitem [{\citenamefont {Okada}\ \emph {et~al.}(2012)\citenamefont {Okada},
  \citenamefont {Zhou}, \citenamefont {Dhital}, \citenamefont {Walkup},
  \citenamefont {Ran}, \citenamefont {Wang}, \citenamefont {Wilson},\ and\
  \citenamefont {Madhavan}}]{PhysRevLett.109.166407}%
  \BibitemOpen
  \bibfield  {author} {\bibinfo {author} {\bibfnamefont {Y.}~\bibnamefont
  {Okada}}, \bibinfo {author} {\bibfnamefont {W.}~\bibnamefont {Zhou}},
  \bibinfo {author} {\bibfnamefont {C.}~\bibnamefont {Dhital}}, \bibinfo
  {author} {\bibfnamefont {D.}~\bibnamefont {Walkup}}, \bibinfo {author}
  {\bibfnamefont {Y.}~\bibnamefont {Ran}}, \bibinfo {author} {\bibfnamefont
  {Z.}~\bibnamefont {Wang}}, \bibinfo {author} {\bibfnamefont {S.~D.}\
  \bibnamefont {Wilson}},\ and\ \bibinfo {author} {\bibfnamefont
  {V.}~\bibnamefont {Madhavan}},\ }\bibfield  {title} {\bibinfo {title}
  {Visualizing $\mathrm{L}$andau $\mathrm{L}$evels of $\mathrm{D}$irac
  $\mathrm{E}$lectrons in a $\mathrm{O}$ne-$\mathrm{D}$imensional
  $\mathrm{P}$otential},\ }\href
  {https://doi.org/10.1103/PhysRevLett.109.166407} {\bibfield  {journal}
  {\bibinfo  {journal} {Phys. Rev. Lett.}\ }\textbf {\bibinfo {volume} {109}},\
  \bibinfo {pages} {166407} (\bibinfo {year} {2012})}\BibitemShut {NoStop}%
\bibitem [{\citenamefont {Sessi}\ \emph {et~al.}(2013)\citenamefont {Sessi},
  \citenamefont {Otrokov}, \citenamefont {Bathon}, \citenamefont {Vergniory},
  \citenamefont {Tsirkin}, \citenamefont {Kokh}, \citenamefont {Tereshchenko},
  \citenamefont {Chulkov},\ and\ \citenamefont {Bode}}]{PhysRevB.88.161407}%
  \BibitemOpen
  \bibfield  {author} {\bibinfo {author} {\bibfnamefont {P.}~\bibnamefont
  {Sessi}}, \bibinfo {author} {\bibfnamefont {M.~M.}\ \bibnamefont {Otrokov}},
  \bibinfo {author} {\bibfnamefont {T.}~\bibnamefont {Bathon}}, \bibinfo
  {author} {\bibfnamefont {M.~G.}\ \bibnamefont {Vergniory}}, \bibinfo {author}
  {\bibfnamefont {S.~S.}\ \bibnamefont {Tsirkin}}, \bibinfo {author}
  {\bibfnamefont {K.~A.}\ \bibnamefont {Kokh}}, \bibinfo {author}
  {\bibfnamefont {O.~E.}\ \bibnamefont {Tereshchenko}}, \bibinfo {author}
  {\bibfnamefont {E.~V.}\ \bibnamefont {Chulkov}},\ and\ \bibinfo {author}
  {\bibfnamefont {M.}~\bibnamefont {Bode}},\ }\bibfield  {title} {\bibinfo
  {title} {Visualizing spin-dependent bulk scattering and breakdown of the
  linear dispersion relation in
  $\mathrm{B}{\mathrm{i}}_{2}\mathrm{T}{\mathrm{e}}_{3}$},\ }\href
  {https://doi.org/10.1103/PhysRevB.88.161407} {\bibfield  {journal} {\bibinfo
  {journal} {Phys. Rev. B}\ }\textbf {\bibinfo {volume} {88}},\ \bibinfo
  {pages} {161407} (\bibinfo {year} {2013})}\BibitemShut {NoStop}%
\bibitem [{\citenamefont {Sessi}\ \emph {et~al.}(2014)\citenamefont {Sessi},
  \citenamefont {Reis}, \citenamefont {Bathon}, \citenamefont {Kokh},
  \citenamefont {Tereshchenko},\ and\ \citenamefont
  {Bode}}]{sessi2014signatures}%
  \BibitemOpen
  \bibfield  {author} {\bibinfo {author} {\bibfnamefont {P.}~\bibnamefont
  {Sessi}}, \bibinfo {author} {\bibfnamefont {F.}~\bibnamefont {Reis}},
  \bibinfo {author} {\bibfnamefont {T.}~\bibnamefont {Bathon}}, \bibinfo
  {author} {\bibfnamefont {K.~A.}\ \bibnamefont {Kokh}}, \bibinfo {author}
  {\bibfnamefont {O.~E.}\ \bibnamefont {Tereshchenko}},\ and\ \bibinfo {author}
  {\bibfnamefont {M.}~\bibnamefont {Bode}},\ }\bibfield  {title} {\bibinfo
  {title} {Signatures of $\mathrm{D}$irac fermion-mediated magnetic order},\
  }\href {https://doi.org/10.1038/ncomms6349} {\bibfield  {journal} {\bibinfo
  {journal} {Nature Communications}\ }\textbf {\bibinfo {volume} {5}},\
  \bibinfo {pages} {5349} (\bibinfo {year} {2014})}\BibitemShut {NoStop}%
\bibitem [{\citenamefont {Bathon}\ \emph {et~al.}(2016)\citenamefont {Bathon},
  \citenamefont {Achilli}, \citenamefont {Sessi}, \citenamefont {Golyashov},
  \citenamefont {Kokh}, \citenamefont {Tereshchenko},\ and\ \citenamefont
  {Bode}}]{https://doi.org/10.1002/adma.201504771}%
  \BibitemOpen
  \bibfield  {author} {\bibinfo {author} {\bibfnamefont {T.}~\bibnamefont
  {Bathon}}, \bibinfo {author} {\bibfnamefont {S.}~\bibnamefont {Achilli}},
  \bibinfo {author} {\bibfnamefont {P.}~\bibnamefont {Sessi}}, \bibinfo
  {author} {\bibfnamefont {V.~A.}\ \bibnamefont {Golyashov}}, \bibinfo {author}
  {\bibfnamefont {K.~A.}\ \bibnamefont {Kokh}}, \bibinfo {author}
  {\bibfnamefont {O.~E.}\ \bibnamefont {Tereshchenko}},\ and\ \bibinfo {author}
  {\bibfnamefont {M.}~\bibnamefont {Bode}},\ }\bibfield  {title} {\bibinfo
  {title} {Experimental $\mathrm{R}$ealization of a $\mathrm{T}$opological
  p–n $\mathrm{J}$unction by $\mathrm{I}$ntrinsic $\mathrm{D}$efect
  $\mathrm{G}$rading},\ }\href
  {https://doi.org/https://doi.org/10.1002/adma.201504771} {\bibfield
  {journal} {\bibinfo  {journal} {Advanced Materials}\ }\textbf {\bibinfo
  {volume} {28}},\ \bibinfo {pages} {2183} (\bibinfo {year}
  {2016})}\BibitemShut {NoStop}%
\bibitem [{\citenamefont {Netsou}\ \emph {et~al.}(2020)\citenamefont {Netsou},
  \citenamefont {Muzychenko}, \citenamefont {Dausy}, \citenamefont {Chen},
  \citenamefont {Song}, \citenamefont {Schouteden}, \citenamefont {Van~Bael},\
  and\ \citenamefont {Van~Haesendonck}}]{netsou2020identifying}%
  \BibitemOpen
  \bibfield  {author} {\bibinfo {author} {\bibfnamefont {A.-M.}\ \bibnamefont
  {Netsou}}, \bibinfo {author} {\bibfnamefont {D.~A.}\ \bibnamefont
  {Muzychenko}}, \bibinfo {author} {\bibfnamefont {H.}~\bibnamefont {Dausy}},
  \bibinfo {author} {\bibfnamefont {T.}~\bibnamefont {Chen}}, \bibinfo {author}
  {\bibfnamefont {F.}~\bibnamefont {Song}}, \bibinfo {author} {\bibfnamefont
  {K.}~\bibnamefont {Schouteden}}, \bibinfo {author} {\bibfnamefont {M.~J.}\
  \bibnamefont {Van~Bael}},\ and\ \bibinfo {author} {\bibfnamefont
  {C.}~\bibnamefont {Van~Haesendonck}},\ }\bibfield  {title} {\bibinfo {title}
  {Identifying $\mathrm{N}$ative $\mathrm{P}$oint $\mathrm{D}$efects in the
  $\mathrm{T}$opological $\mathrm{I}$nsulator
  $\mathrm{B}{\mathrm{i}}_{2}\mathrm{T}{\mathrm{e}}_{3}$},\ }\href
  {https://doi.org/10.1021/acsnano.0c04861} {\bibfield  {journal} {\bibinfo
  {journal} {ACS Nano}\ }\textbf {\bibinfo {volume} {14}},\ \bibinfo {pages}
  {13172} (\bibinfo {year} {2020})}\BibitemShut {NoStop}%
\bibitem [{\citenamefont {Stolyarov}\ \emph {et~al.}(2021)\citenamefont
  {Stolyarov}, \citenamefont {Sheina}, \citenamefont {Khokhlov}, \citenamefont
  {Vlaic}, \citenamefont {Pons}, \citenamefont {Aubin}, \citenamefont
  {Akzyanov}, \citenamefont {Vasenko}, \citenamefont {Menshchikova},
  \citenamefont {Chulkov}, \citenamefont {Golubov}, \citenamefont {Cren},\ and\
  \citenamefont {Roditchev}}]{Stolyarov2021}%
  \BibitemOpen
  \bibfield  {author} {\bibinfo {author} {\bibfnamefont {V.~S.}\ \bibnamefont
  {Stolyarov}}, \bibinfo {author} {\bibfnamefont {V.~A.}\ \bibnamefont
  {Sheina}}, \bibinfo {author} {\bibfnamefont {D.~A.}\ \bibnamefont
  {Khokhlov}}, \bibinfo {author} {\bibfnamefont {S.}~\bibnamefont {Vlaic}},
  \bibinfo {author} {\bibfnamefont {S.}~\bibnamefont {Pons}}, \bibinfo {author}
  {\bibfnamefont {H.}~\bibnamefont {Aubin}}, \bibinfo {author} {\bibfnamefont
  {R.~S.}\ \bibnamefont {Akzyanov}}, \bibinfo {author} {\bibfnamefont {A.~S.}\
  \bibnamefont {Vasenko}}, \bibinfo {author} {\bibfnamefont {T.~V.}\
  \bibnamefont {Menshchikova}}, \bibinfo {author} {\bibfnamefont {E.~V.}\
  \bibnamefont {Chulkov}}, \bibinfo {author} {\bibfnamefont {A.~A.}\
  \bibnamefont {Golubov}}, \bibinfo {author} {\bibfnamefont {T.}~\bibnamefont
  {Cren}},\ and\ \bibinfo {author} {\bibfnamefont {D.}~\bibnamefont
  {Roditchev}},\ }\bibfield  {title} {\bibinfo {title}
  {Disorder-$\mathrm{P}$romoted $\mathrm{S}$plitting in
  $\mathrm{Q}$uasiparticle $\mathrm{I}$nterference at $\mathrm{N}$esting
  $\mathrm{V}$ectors},\ }\href {https://doi.org/10.1021/acs.jpclett.1c00462}
  {\bibfield  {journal} {\bibinfo  {journal} {The Journal of Physical Chemistry
  Letters}\ }\textbf {\bibinfo {volume} {12}},\ \bibinfo {pages} {3127}
  (\bibinfo {year} {2021})}\BibitemShut {NoStop}%
\bibitem [{\citenamefont {Hsieh}\ \emph {et~al.}(2009)\citenamefont {Hsieh},
  \citenamefont {Xia}, \citenamefont {Qian}, \citenamefont {Wray},
  \citenamefont {Dil}, \citenamefont {Meier}, \citenamefont {Osterwalder},
  \citenamefont {Patthey}, \citenamefont {Checkelsky}, \citenamefont {Ong},
  \citenamefont {Fedorov}, \citenamefont {Lin}, \citenamefont {Bansil},
  \citenamefont {Grauer}, \citenamefont {Hor}, \citenamefont {Cava},\ and\
  \citenamefont {Hasan}}]{Hsieh2009}%
  \BibitemOpen
  \bibfield  {author} {\bibinfo {author} {\bibfnamefont {D.}~\bibnamefont
  {Hsieh}}, \bibinfo {author} {\bibfnamefont {Y.}~\bibnamefont {Xia}}, \bibinfo
  {author} {\bibfnamefont {D.}~\bibnamefont {Qian}}, \bibinfo {author}
  {\bibfnamefont {L.}~\bibnamefont {Wray}}, \bibinfo {author} {\bibfnamefont
  {J.~H.}\ \bibnamefont {Dil}}, \bibinfo {author} {\bibfnamefont
  {F.}~\bibnamefont {Meier}}, \bibinfo {author} {\bibfnamefont
  {J.}~\bibnamefont {Osterwalder}}, \bibinfo {author} {\bibfnamefont
  {L.}~\bibnamefont {Patthey}}, \bibinfo {author} {\bibfnamefont {J.~G.}\
  \bibnamefont {Checkelsky}}, \bibinfo {author} {\bibfnamefont {N.~P.}\
  \bibnamefont {Ong}}, \bibinfo {author} {\bibfnamefont {A.~V.}\ \bibnamefont
  {Fedorov}}, \bibinfo {author} {\bibfnamefont {H.}~\bibnamefont {Lin}},
  \bibinfo {author} {\bibfnamefont {A.}~\bibnamefont {Bansil}}, \bibinfo
  {author} {\bibfnamefont {D.}~\bibnamefont {Grauer}}, \bibinfo {author}
  {\bibfnamefont {Y.~S.}\ \bibnamefont {Hor}}, \bibinfo {author} {\bibfnamefont
  {R.~J.}\ \bibnamefont {Cava}},\ and\ \bibinfo {author} {\bibfnamefont
  {M.~Z.}\ \bibnamefont {Hasan}},\ }\bibfield  {title} {\bibinfo {title} {A
  tunable topological insulator in the spin helical $\mathrm{D}$irac transport
  regime},\ }\href {https://doi.org/10.1038/nature08234} {\bibfield  {journal}
  {\bibinfo  {journal} {Nature}\ }\textbf {\bibinfo {volume} {460}},\ \bibinfo
  {pages} {1101} (\bibinfo {year} {2009})}\BibitemShut {NoStop}%
\bibitem [{\citenamefont {Souma}\ \emph {et~al.}(2011)\citenamefont {Souma},
  \citenamefont {Kosaka}, \citenamefont {Sato}, \citenamefont {Komatsu},
  \citenamefont {Takayama}, \citenamefont {Takahashi}, \citenamefont {Kriener},
  \citenamefont {Segawa},\ and\ \citenamefont {Ando}}]{PhysRevLett.106.216803}%
  \BibitemOpen
  \bibfield  {author} {\bibinfo {author} {\bibfnamefont {S.}~\bibnamefont
  {Souma}}, \bibinfo {author} {\bibfnamefont {K.}~\bibnamefont {Kosaka}},
  \bibinfo {author} {\bibfnamefont {T.}~\bibnamefont {Sato}}, \bibinfo {author}
  {\bibfnamefont {M.}~\bibnamefont {Komatsu}}, \bibinfo {author} {\bibfnamefont
  {A.}~\bibnamefont {Takayama}}, \bibinfo {author} {\bibfnamefont
  {T.}~\bibnamefont {Takahashi}}, \bibinfo {author} {\bibfnamefont
  {M.}~\bibnamefont {Kriener}}, \bibinfo {author} {\bibfnamefont
  {K.}~\bibnamefont {Segawa}},\ and\ \bibinfo {author} {\bibfnamefont
  {Y.}~\bibnamefont {Ando}},\ }\bibfield  {title} {\bibinfo {title} {Direct
  $\mathrm{M}$easurement of the $\mathrm{O}$ut-of-$\mathrm{P}$lane
  $\mathrm{S}$pin $\mathrm{T}$exture in the $\mathrm{D}$irac-$\mathrm{C}$one
  $\mathrm{S}$urface $\mathrm{S}$tate of a $\mathrm{T}$opological
  $\mathrm{I}$nsulator},\ }\href
  {https://doi.org/10.1103/PhysRevLett.106.216803} {\bibfield  {journal}
  {\bibinfo  {journal} {Phys. Rev. Lett.}\ }\textbf {\bibinfo {volume} {106}},\
  \bibinfo {pages} {216803} (\bibinfo {year} {2011})}\BibitemShut {NoStop}%
\bibitem [{\citenamefont {Schlegel}\ \emph {et~al.}(2014)\citenamefont
  {Schlegel}, \citenamefont {Hänke}, \citenamefont {Baumann}, \citenamefont
  {Kaiser}, \citenamefont {Nag}, \citenamefont {Voigtländer}, \citenamefont
  {Lindackers}, \citenamefont {Büchner},\ and\ \citenamefont
  {Hess}}]{Schlegel2014}%
  \BibitemOpen
  \bibfield  {author} {\bibinfo {author} {\bibfnamefont {R.}~\bibnamefont
  {Schlegel}}, \bibinfo {author} {\bibfnamefont {T.}~\bibnamefont {Hänke}},
  \bibinfo {author} {\bibfnamefont {D.}~\bibnamefont {Baumann}}, \bibinfo
  {author} {\bibfnamefont {M.}~\bibnamefont {Kaiser}}, \bibinfo {author}
  {\bibfnamefont {P.~K.}\ \bibnamefont {Nag}}, \bibinfo {author} {\bibfnamefont
  {R.}~\bibnamefont {Voigtländer}}, \bibinfo {author} {\bibfnamefont
  {D.}~\bibnamefont {Lindackers}}, \bibinfo {author} {\bibfnamefont
  {B.}~\bibnamefont {Büchner}},\ and\ \bibinfo {author} {\bibfnamefont
  {C.}~\bibnamefont {Hess}},\ }\bibfield  {title} {\bibinfo {title} {Design and
  properties of a cryogenic dip-stick scanning tunneling microscope with
  capacitive coarse approach control},\ }\href
  {https://doi.org/10.1063/1.4862817} {\bibfield  {journal} {\bibinfo
  {journal} {Review of Scientific Instruments}\ }\textbf {\bibinfo {volume}
  {85}},\ \bibinfo {pages} {013706} (\bibinfo {year} {2014})}\BibitemShut
  {NoStop}%
\bibitem [{Spe()}]{Specs}%
  \BibitemOpen
  \href@noop {} {\emph {\bibinfo {title} {Specs, Nanonis SPM Controller, SPECS
  Surface Nano Analysis GmbH, Voltastraße 5, D-13355 Berlin,
  Germany.}}}\BibitemShut {Stop}%
\bibitem [{\citenamefont {Horcas}\ \emph {et~al.}(2007)\citenamefont {Horcas},
  \citenamefont {Fernández}, \citenamefont {Gómez-Rodríguez}, \citenamefont
  {Colchero}, \citenamefont {Gómez-Herrero},\ and\ \citenamefont
  {Baro}}]{Horcas2007}%
  \BibitemOpen
  \bibfield  {author} {\bibinfo {author} {\bibfnamefont {I.}~\bibnamefont
  {Horcas}}, \bibinfo {author} {\bibfnamefont {R.}~\bibnamefont {Fernández}},
  \bibinfo {author} {\bibfnamefont {J.~M.}\ \bibnamefont {Gómez-Rodríguez}},
  \bibinfo {author} {\bibfnamefont {J.}~\bibnamefont {Colchero}}, \bibinfo
  {author} {\bibfnamefont {J.}~\bibnamefont {Gómez-Herrero}},\ and\ \bibinfo
  {author} {\bibfnamefont {A.~M.}\ \bibnamefont {Baro}},\ }\bibfield  {title}
  {\bibinfo {title} {W$\mathrm{S}$x$\mathrm{M}$: A software for scanning probe
  microscopy and a tool for nanotechnology},\ }\href
  {https://doi.org/10.1063/1.2432410} {\bibfield  {journal} {\bibinfo
  {journal} {Review of Scientific Instruments}\ }\textbf {\bibinfo {volume}
  {78}},\ \bibinfo {pages} {013705} (\bibinfo {year} {2007})}\BibitemShut
  {NoStop}%
\bibitem [{\citenamefont {Beidenkopf}\ \emph {et~al.}(2011)\citenamefont
  {Beidenkopf}, \citenamefont {Roushan}, \citenamefont {Seo}, \citenamefont
  {Gorman}, \citenamefont {Drozdov}, \citenamefont {Hor}, \citenamefont
  {Cava},\ and\ \citenamefont {Yazdani}}]{Beidenkopf2011}%
  \BibitemOpen
  \bibfield  {author} {\bibinfo {author} {\bibfnamefont {H.}~\bibnamefont
  {Beidenkopf}}, \bibinfo {author} {\bibfnamefont {P.}~\bibnamefont {Roushan}},
  \bibinfo {author} {\bibfnamefont {J.}~\bibnamefont {Seo}}, \bibinfo {author}
  {\bibfnamefont {L.}~\bibnamefont {Gorman}}, \bibinfo {author} {\bibfnamefont
  {I.}~\bibnamefont {Drozdov}}, \bibinfo {author} {\bibfnamefont {Y.~S.}\
  \bibnamefont {Hor}}, \bibinfo {author} {\bibfnamefont {R.~J.}\ \bibnamefont
  {Cava}},\ and\ \bibinfo {author} {\bibfnamefont {A.}~\bibnamefont
  {Yazdani}},\ }\bibfield  {title} {\bibinfo {title} {Spatial fluctuations of
  helical $\mathrm{D}$irac fermions on the surface of topological insulators},\
  }\href {https://doi.org/10.1038/nphys2108} {\bibfield  {journal} {\bibinfo
  {journal} {Nature Physics}\ }\textbf {\bibinfo {volume} {7}},\ \bibinfo
  {pages} {939} (\bibinfo {year} {2011})}\BibitemShut {NoStop}%
\bibitem [{\citenamefont {Lee}\ \emph {et~al.}(2009)\citenamefont {Lee},
  \citenamefont {Wu}, \citenamefont {Arovas},\ and\ \citenamefont
  {Zhang}}]{PhysRevB.80.245439}%
  \BibitemOpen
  \bibfield  {author} {\bibinfo {author} {\bibfnamefont {W.-C.}\ \bibnamefont
  {Lee}}, \bibinfo {author} {\bibfnamefont {C.}~\bibnamefont {Wu}}, \bibinfo
  {author} {\bibfnamefont {D.~P.}\ \bibnamefont {Arovas}},\ and\ \bibinfo
  {author} {\bibfnamefont {S.-C.}\ \bibnamefont {Zhang}},\ }\bibfield  {title}
  {\bibinfo {title} {Quasiparticle interference on the surface of the
  topological insulator
  $\mathrm{B}{\mathrm{i}}_{2}\mathrm{T}{\mathrm{e}}_{3}$},\ }\href
  {https://doi.org/10.1103/PhysRevB.80.245439} {\bibfield  {journal} {\bibinfo
  {journal} {Phys. Rev. B}\ }\textbf {\bibinfo {volume} {80}},\ \bibinfo
  {pages} {245439} (\bibinfo {year} {2009})}\BibitemShut {NoStop}%
\bibitem [{\citenamefont {Kim}\ \emph {et~al.}(2011)\citenamefont {Kim},
  \citenamefont {Ye}, \citenamefont {Kuroda}, \citenamefont {Yamada},
  \citenamefont {Krasovskii}, \citenamefont {Chulkov}, \citenamefont
  {Miyamoto}, \citenamefont {Nakatake}, \citenamefont {Okuda}, \citenamefont
  {Ueda}, \citenamefont {Shimada}, \citenamefont {Namatame}, \citenamefont
  {Taniguchi},\ and\ \citenamefont {Kimura}}]{PhysRevLett.107.056803}%
  \BibitemOpen
  \bibfield  {author} {\bibinfo {author} {\bibfnamefont {S.}~\bibnamefont
  {Kim}}, \bibinfo {author} {\bibfnamefont {M.}~\bibnamefont {Ye}}, \bibinfo
  {author} {\bibfnamefont {K.}~\bibnamefont {Kuroda}}, \bibinfo {author}
  {\bibfnamefont {Y.}~\bibnamefont {Yamada}}, \bibinfo {author} {\bibfnamefont
  {E.~E.}\ \bibnamefont {Krasovskii}}, \bibinfo {author} {\bibfnamefont
  {E.~V.}\ \bibnamefont {Chulkov}}, \bibinfo {author} {\bibfnamefont
  {K.}~\bibnamefont {Miyamoto}}, \bibinfo {author} {\bibfnamefont
  {M.}~\bibnamefont {Nakatake}}, \bibinfo {author} {\bibfnamefont
  {T.}~\bibnamefont {Okuda}}, \bibinfo {author} {\bibfnamefont
  {Y.}~\bibnamefont {Ueda}}, \bibinfo {author} {\bibfnamefont {K.}~\bibnamefont
  {Shimada}}, \bibinfo {author} {\bibfnamefont {H.}~\bibnamefont {Namatame}},
  \bibinfo {author} {\bibfnamefont {M.}~\bibnamefont {Taniguchi}},\ and\
  \bibinfo {author} {\bibfnamefont {A.}~\bibnamefont {Kimura}},\ }\bibfield
  {title} {\bibinfo {title} {Surface $\mathrm{S}$cattering via $\mathrm{B}$ulk
  $\mathrm{C}$ontinuum $\mathrm{S}$tates in the 3$\mathrm{D}$
  $\mathrm{T}$opological $\mathrm{I}$nsulator
  $\mathrm{B}{\mathrm{i}}_{2}\mathrm{S}{\mathrm{e}}_{3}$},\ }\href
  {https://doi.org/10.1103/PhysRevLett.107.056803} {\bibfield  {journal}
  {\bibinfo  {journal} {Phys. Rev. Lett.}\ }\textbf {\bibinfo {volume} {107}},\
  \bibinfo {pages} {056803} (\bibinfo {year} {2011})}\BibitemShut {NoStop}%
\bibitem [{\citenamefont {Hedayat}\ \emph {et~al.}(2021)\citenamefont
  {Hedayat}, \citenamefont {Bugini}, \citenamefont {Yi}, \citenamefont {Chen},
  \citenamefont {Zhou}, \citenamefont {Cerullo}, \citenamefont {Dallera},\ and\
  \citenamefont {Carpene}}]{Hedayat2021}%
  \BibitemOpen
  \bibfield  {author} {\bibinfo {author} {\bibfnamefont {H.}~\bibnamefont
  {Hedayat}}, \bibinfo {author} {\bibfnamefont {D.}~\bibnamefont {Bugini}},
  \bibinfo {author} {\bibfnamefont {H.}~\bibnamefont {Yi}}, \bibinfo {author}
  {\bibfnamefont {C.}~\bibnamefont {Chen}}, \bibinfo {author} {\bibfnamefont
  {X.}~\bibnamefont {Zhou}}, \bibinfo {author} {\bibfnamefont {G.}~\bibnamefont
  {Cerullo}}, \bibinfo {author} {\bibfnamefont {C.}~\bibnamefont {Dallera}},\
  and\ \bibinfo {author} {\bibfnamefont {E.}~\bibnamefont {Carpene}},\
  }\bibfield  {title} {\bibinfo {title} {Ultrafast evolution of bulk, surface
  and surface resonance states in photoexcited
  $\mathrm{B}{\mathrm{i}}_{2}\mathrm{T}{\mathrm{e}}_{3}$},\ }\href
  {https://doi.org/10.1038/s41598-021-83848-z} {\bibfield  {journal} {\bibinfo
  {journal} {Scientific Reports}\ }\textbf {\bibinfo {volume} {11}},\ \bibinfo
  {pages} {4924} (\bibinfo {year} {2021})}\BibitemShut {NoStop}%
\bibitem [{\citenamefont {Hsu}\ \emph {et~al.}(2014)\citenamefont {Hsu},
  \citenamefont {Fischer}, \citenamefont {Hughes}, \citenamefont {Park},\ and\
  \citenamefont {Kim}}]{Hsu2014}%
  \BibitemOpen
  \bibfield  {author} {\bibinfo {author} {\bibfnamefont {Y.-T.}\ \bibnamefont
  {Hsu}}, \bibinfo {author} {\bibfnamefont {M.~H.}\ \bibnamefont {Fischer}},
  \bibinfo {author} {\bibfnamefont {T.~L.}\ \bibnamefont {Hughes}}, \bibinfo
  {author} {\bibfnamefont {K.}~\bibnamefont {Park}},\ and\ \bibinfo {author}
  {\bibfnamefont {E.-A.}\ \bibnamefont {Kim}},\ }\bibfield  {title} {\bibinfo
  {title} {Effects of surface-bulk hybridization in three-dimensional
  topological metals},\ }\href {https://doi.org/10.1103/PhysRevB.89.205438}
  {\bibfield  {journal} {\bibinfo  {journal} {Phys. Rev. B}\ }\textbf {\bibinfo
  {volume} {89}},\ \bibinfo {pages} {205438} (\bibinfo {year}
  {2014})}\BibitemShut {NoStop}%
\bibitem [{Foo()}]{Footnote}%
  \BibitemOpen
  \href@noop {} {\bibinfo {title} {$\mathrm{W}$e mention that a double
  $\mathrm{QPI}$ structure in the $\mathrm{\Gamma}$-$\mathrm{M}$ direction at
  different energies has also been recently reported in $\mathrm{R}$ef. 26.
  $\mathrm{T}$here it was observed at a completely different energy range,
  where a significant warping of the $\mathrm{D}$irac $\mathrm{CEC}$ is known,
  and the splitting naturally results from that warping. $\mathrm{N}$ote that
  in our considered energy range, the warping, according to $\mathrm{ARPES}$
  \cite{doi:10.1126/science.1173034, PhysRevLett.104.016401}, is practically
  absent. $\mathrm{T}$hus, our observed splitting must be of different
  origin}}\BibitemShut {NoStop}%
\bibitem [{\citenamefont {Wittel}\ and\ \citenamefont
  {Manne}(1974)}]{Wittel1974}%
  \BibitemOpen
  \bibfield  {author} {\bibinfo {author} {\bibfnamefont {K.}~\bibnamefont
  {Wittel}}\ and\ \bibinfo {author} {\bibfnamefont {R.}~\bibnamefont {Manne}},\
  }\bibfield  {title} {\bibinfo {title} {Atomic
  $\mathrm{S}$pin-$\mathrm{O}$rbit $\mathrm{I}$nteraction $\mathrm{P}$arameters
  from $\mathrm{S}$pectral $\mathrm{D}$ata for 19 $\mathrm{E}$lements},\ }\href
  {https://doi.org/10.1007/BF00551162} {\bibfield  {journal} {\bibinfo
  {journal} {Theoretica chimica acta}\ }\textbf {\bibinfo {volume} {33}},\
  \bibinfo {pages} {347} (\bibinfo {year} {1974})}\BibitemShut {NoStop}%
\bibitem [{\citenamefont {Shallenberger}\ \emph {et~al.}(2019)\citenamefont
  {Shallenberger}, \citenamefont {Smyth}, \citenamefont {Addou},\ and\
  \citenamefont {Wallace}}]{Shallenberger2019}%
  \BibitemOpen
  \bibfield  {author} {\bibinfo {author} {\bibfnamefont {J.~R.}\ \bibnamefont
  {Shallenberger}}, \bibinfo {author} {\bibfnamefont {C.~M.}\ \bibnamefont
  {Smyth}}, \bibinfo {author} {\bibfnamefont {R.}~\bibnamefont {Addou}},\ and\
  \bibinfo {author} {\bibfnamefont {R.~M.}\ \bibnamefont {Wallace}},\
  }\bibfield  {title} {\bibinfo {title} {2$\mathrm{D}$ bismuth telluride
  analyzed by $\mathrm{XPS}$},\ }\href {https://doi.org/10.1116/1.5120015}
  {\bibfield  {journal} {\bibinfo  {journal} {Surface Science Spectra}\
  }\textbf {\bibinfo {volume} {26}},\ \bibinfo {pages} {024011} (\bibinfo
  {year} {2019})}\BibitemShut {NoStop}%
\bibitem [{\citenamefont {Concepción}\ \emph {et~al.}(2018)\citenamefont
  {Concepción}, \citenamefont {Galván-Arellano}, \citenamefont
  {Torres-Costa}, \citenamefont {Climent-Font}, \citenamefont {Bahena},
  \citenamefont {Manso~Silván}, \citenamefont {Escobosa},\ and\ \citenamefont
  {de~Melo}}]{Concepcion2018}%
  \BibitemOpen
  \bibfield  {author} {\bibinfo {author} {\bibfnamefont {O.}~\bibnamefont
  {Concepción}}, \bibinfo {author} {\bibfnamefont {M.}~\bibnamefont
  {Galván-Arellano}}, \bibinfo {author} {\bibfnamefont {V.}~\bibnamefont
  {Torres-Costa}}, \bibinfo {author} {\bibfnamefont {A.}~\bibnamefont
  {Climent-Font}}, \bibinfo {author} {\bibfnamefont {D.}~\bibnamefont
  {Bahena}}, \bibinfo {author} {\bibfnamefont {M.}~\bibnamefont
  {Manso~Silván}}, \bibinfo {author} {\bibfnamefont {A.}~\bibnamefont
  {Escobosa}},\ and\ \bibinfo {author} {\bibfnamefont {O.}~\bibnamefont
  {de~Melo}},\ }\bibfield  {title} {\bibinfo {title} {Controlling the
  $\mathrm{E}$pitaxial $\mathrm{G}$rowth of
  $\mathrm{B}{\mathrm{i}}_{2}\mathrm{T}{\mathrm{e}}_{3}$,
  $\mathrm{B}$i$\mathrm{T}$e, and
  $\mathrm{B}{\mathrm{i}}_{4}\mathrm{T}{\mathrm{e}}_{3}$ $\mathrm{P}$ure
  $\mathrm{P}$hases by $\mathrm{P}$hysical $\mathrm{V}$apor
  $\mathrm{T}$ransport},\ }\href
  {https://doi.org/10.1021/acs.inorgchem.8b01235} {\bibfield  {journal}
  {\bibinfo  {journal} {Inorganic Chemistry}\ }\textbf {\bibinfo {volume}
  {57}},\ \bibinfo {pages} {10090} (\bibinfo {year} {2018})}\BibitemShut
  {NoStop}%
\end{thebibliography}%
\end{document}